\documentstyle [12pt]{article}
\begin{document}
\newcommand{\vm}{\vspace{0.2cm}}
\newcommand{\vl}{\vspace{0.4cm}}
\vl

\title{p-Adic  description of Higgs mechanism IV: elementary particle  and
 hadron masses }
\author{Matti Pitk\"anen\\
Torkkelinkatu 21 B 39, 00530, Helsinki, FINLAND}
\date{6.10. 1994}
\maketitle

\newpage

\begin{center}
Abstract
\end{center}

\vl

This paper is the fourth  one  in the series devoted to  the calculation
of particle masses in the framework of p-adic
 conformal field theory limit of Topological GeometroDynamics. In the third
paper the masses of  elementary fermions and bosons were calculated and
 in this paper these results are applied. The masses of  charged leptons
and $W$ boson
 are predicted with accuracy better than one per cent. Weinberg angle is
predicted to be $sin^2(\theta_W)=3/8$ so that $Z^0$ mass is $10$ per cent
too large. One can reproduce lepton and gauge boson masses  correctly by
taking into account Coulombic self energy associated with the interior of
3-surface and a small
 mixing of boundary topologies associated with charged leptons.   The
general mass formula for hadrons involves
 boundary contributions of quarks calculated in previous paper plus
interior  term consisting
 isospi-isopin,  color magnetic spin-spin and color Coulombic terms. One
must take also into account
 topological mixing of boundary topologies plus the mixing for primary
condensate levels  (quark spends part of time at  lower condensate
level).  Topological mixing implies the nontriviality of
 CKM matrix. If (the moduli
  squared of)   mixing  matrix elements are rational numbers,  mixing
parameters must  satisfy strong number theoretical constraints. It is
possible to reproduce CKM matrix satisfying the empirical constraints  but
an open question is whether the number theoretic conditions can be
satisfied.   CP breaking is a  number theoretical necessity. The mixing of
$u$ quark with  $c$ quark  is
 large and solves  the spin crisis of proton. The parameters associated
with the
   interior $O(p)$ contribution  can be fixed  by   no Planck mass
condition plus some empirical inputs.   $O(p)$ contribution to mass
dominates for all hadrons except pion for which  $O(p^2)$ contribution
from  quark masses predicts mass correctly within few per cent. $O(p^2)$
contribution determines
 isospin  splitting:
 besides quark masses
 Coulomb interaction contributes to  isospin splitting. In p-adic case
also  constant shifts coming from color magnetic spin- spin interaction,
from isospin-isospin interaction,
  from the masses  of
 sea partons,.. are important since modular mathematics is involved.
Observed splittings can
 be reproduced but only few predictions (correct within experimental
limits) are possible.  The observed masses are  reproduced with at most
one per cent error.
 The only exception is top quark,  whose mass  is predicted to be  about
$5$  times larger (3 times smaller)  than the mass of observed top candidate
for
$k=89$ ($k=97$).  Arguments related to CP breaking seem to exclude
$k=89$ and $k=97$ condensation levels.  A rather plausible resolution of the
problem is small mixing of $k=97$
 and $k=89$ condensate levels implying increase in the mass of the top.  An
alternative possibility is that the observed top candidate in fact
corresponds to lowest
 generation  ($g=0$) quark of $M_{89}$ physics obtained by scaling up the
ordinary $M_{107}$  physics by the ratio  $\sqrt{M_{107}/M_{89}}\simeq
512$ of Mersenne primes. The resulting prediction for the mass of top
candidate is correct. The third alternative is that mixing model is correct and
$M_{89}$ mesons with nearly same energy as $t\bar{t}$ mesons are also produced:
this might explain  the reported anomalies in  the decay characteristics.

\newpage

\tableofcontents

\newpage

\section{Introduction}

This is the fourth paper in series devoted to the p-adic description of
Higgs mechanism in TGD.   In the first paper the general formulation of
p-adic conformal field theory
 limit of
 TGD was proposed and reader is suggested to read the introduction of this
 paper for general background.  In paper III the calculations of
elementary fermion and boson masses were carried out. In this paper the
results of the calculation  will be analyzed in detail
 and will be extended to a model predicting hadron masses with one per
cent accuracy.

\vm

The predictions of TGD for lepton and gauge bosons masses are correct
 with error smaller than one percent except for $Z^0$ boson, for which
mass is too large by about 10 per cent.  These discrepancies can be
understood as following from the neglect of Coulombic self energy and
small mixing of boundary topologies for leptons and when  these effects
are taken  into account the  masses of leptons  and intermediate gauge
bosons can be reproduced  exactly.    A good guess for renormalized
Weinberg angle comes from the requirement that  both $sin(\theta_W)$ and
$cos(\theta_W)$ are rational numbers and therefore
 correspond to Pythagorean triangle. For minimal Pythagorean triangle one
obtains the prediction  $sin^2(\theta_W)=8^2/17^2\simeq 0.2215$ in
accordance with the value deduced  from $\nu N$ scattering experiments
\cite{Nenucl}.

\vm

 TGD predicts also colored excitations of leptons and quarks. The
existence of  colored leptons is in accordance with the leptopion
hypothesis \cite{Heavy,Lepto} used to explain the
 anomalous $e^+e^-$ production observed in heavy ion collisions. On the
other hand, $Z^0$  decay widths plus the cherished hypothesis of
asymptotic freedom  seem to require these excitations to be very massive
(to condense on condensate level with small $p$).   These issues
 will be considered in the fifth paper of the series.

\vm

 Hadron mass formula involves several contributions. \\ a) The
contribution of valence and sea quarks. Thermodynamical  equilibrium
assumption together with quantization of chemical potentials at low
temperature  limit  implies that sea partons give only  $O(p^2)$
contribution to masses. The masses of quarks   were calculated in previous
paper and mass formulas are almost identical with leptonic mass
formulas.\\ b)  The interior $O(p)$  contribution coming from
  color magnetic spin-spin interaction,  electroweak  isospin-isospin
interaction plus color Coulombic interaction.   The assumption of strong
electrweak  isospin-isospin interaction means deviation from the picture
of standard model. In order $O(p)$ these  contributions are parametrized
by small integers and few empirical inputs plus no Planck mass
 condition fix the values of these integers completely.  \\ c)
Topological mixing of boundary topologies
 must be taken into account and few empirical inputs fix the mixing
scenario essentially uniquely
 predicting Cabibbo angle  correctly. The requirement that the moduli
squared  for   U and D  and
 CKM matrix elements are rational numbers implies strong number
theoretical conditions on  the parameters appearing in the mixing matrices
for U and D type quarks and it is an open problem whether
 these conditions can be satisfied. An even stronger condition is the
rationality of U,D and CKM  matrices.   CKM matrix satisfying empirical
constraints can be found with essentially unique values of
 mixing parameters and a solution of spin crisis of proton follows as a
byproduct. What happens is  that $u$ quark mixes unexpectedly strongly
with c quark so that $g=0$  contribution to the spin of
 proton becomes of order $1/3$.  CP breaking is a number theoretic
necessity and comes out correctly as
 a prediction, when the values of various  parameters are determined  from
the empirical  estimates for  the moduli squared of  CKM matrix elements.
\\
 d) Also the mixing of primary condensate levels is needed in order to
 understand the lowering of the masses of hadrons b or c type quark plus
u,d or s quark.    The  mixing of
 primary condensate level must be assumed:  for instance,   b quark
(level $k=103$ with $p \simeq 2^k$, $k$ prime ) spends fraction of time
 as condensed on
 u,d or s quark (level $k=107$)  and this lowers the effective b quark
mass in hadrons containing u,d or s quarks.    \\  e)  The second order
term in  hadronic  mass   squared  receives contributions
 from
 second order terms in   quark masses, from  the masses of sea partons,
from electromagnetic and color magnetic
  spin-spin interaction,  electroweak isospin-isospin  interaction and
color Coulombic and ordinary  Coulombic interaction.
 Although quark masses and electromagnetic spin-spin interaction  give
the only  contributions   depending  on isospin,   the constant multiplet
shifts  implied by  the other contributions are important in p-adic
context since
 delicate modulo mathematics is involved.  Since so many effects are
involved,   only a few predictions  are possible at this stage. \\
  f)   In general the  predictions are  typically correct within one  per
cent.
 There is only single notable exception.  Top quark condensed on condensate
level $k=89$ ($k=97$)
 is predicted to have mass about $5$ times larger ($3$ times smaller)
than the mass of the observed top candidate.  Arguments related to CP
breaking seem to force the top quark mass to be the mass of the observed
top candidate and small condensate level mixing can indeed explain top mass.
   $g=0$ quarks of  $M_{89}$ hadron physics, which is
just scaled up copy of $M_{107}$
 physics  (predicted to exist in \cite{padTGD}) have however masses quite near
to the mass of  the top candidate and production  of $M_{89}$ hadrons
might explain the reported anomalies in the production and decay of top
candidate.
 The
experimental signatures of new hadron physics will be discussed in the fifth
paper of the series toghether with the
 problems related  to the  possible existence of light  colored
excitations of leptons and quarks.

 \section{Lepton masses }

If interior contributions are neglected fermion mass squared is   sum of cm
 and modular contributions related to the boundary component

\begin{eqnarray}
M^2&=&M^2(cm) +M^2(mod)
\end{eqnarray}

\noindent  'cm'   refers to the cm of boundary component.  cm contribution
is same for all families and depends on isospin of lepton.  'mod'  refers
to the contribution of modular degrees of freedom associated with boundary
component.  There is in principle present also a contribution coming from
the interior of leptonic 3-surface and this contribution can be identified
tentatively as electroweak self interaction energy: this contribution
turns out to be smaller than percent as also suggested by dimensional
considerations.

\vm

 Modular contribution is given by the formula

\begin{eqnarray}
M^2(mod,g)  &=& k(F) 2N(g)g \frac{M^2_0}{p}\nonumber\\
k(F)&=&\frac{3}{2}\nonumber\\
k(mod)&=&1
\end{eqnarray}

\noindent and has no dependence on electroweak and other standard quantum
numbers.

\subsection{Charged lepton masses}

For charged leptons cm  contribution is under rather general assumptions
about
 prime $p$ given by the formula

\begin{eqnarray}
M^2(cm) &=&(5+  \frac{2}{3}) \frac{M^2_0}{p(L)}
\end{eqnarray}

\noindent   The explicit form of the
 mass formula in case of charged leptons is

\begin{eqnarray}
M^2(g,p)&=& (5+\frac{2}{3} +3N(g)g) \frac{M_0^2}{p}\nonumber\\
N(g)&=& 2^{g-1}(2^{g}+1)
\end{eqnarray}

\noindent The  formula gives following prediction for  lepton masses

\begin{eqnarray}
M^2(e)&=& (5+\frac{2}{3})\frac{M_0^2}{M_{127}}\nonumber\\
M^2(\mu)&=& (14+\frac{2}{3})\frac{M_0^2}{p(\mu)}, \ p(\mu) \simeq 2^{113}
\nonumber\\
M^2(\tau)&=& (65+\frac{2}{3}) \frac{M_0^2}{M_{107}}
\end{eqnarray}

\noindent  The  relative errors of lepton masses are below
one percent as the table shows.

\vl

\begin{tabular}{||c|c|c|c||}
 \hline\hline
particle &$ m_{exp}/MeV$ &$m_{pred}/MeV$&rel. error/\%\\ \cline{1-4}\hline
e &0.5110034&0.5110034& 0\\ \hline
$\mu$ &105.659&105.229&-0.4\\ \hline\hline
$\tau$ &1776.9&1781.3&+0.2\\ \hline\hline
$W$ &80200&79582&-0.7\\ \hline\hline
$Z$ &91151&100664&10.0\\ \hline\hline
\end{tabular}

\vl

Table 2.1. \label{Leptmasses} Predictions for masses of charged leptons and
 intermediate gauge bosons.

\vm

W/e  mass ratio is  predicted with smaller than one per cent error  but
$Z^0$ mass is too large by  ten per cent.

\subsection{Effects of Coulombic self interaction and topology mixing
on charged lepton masses}

The predictions for the masses of $\mu$ and $\tau$ as well as $W$ boson are
 too small by less then one per cent. The probable reasons for the
discrepancy are  the neglect of the interior contribution to mass squared
and the mixing of boundary topologies. \\
 a)   The first guess is that interior  contribution must corresponds to
the electroweak self interaction energy of the particle, which is
inversely proportional to p-adic length scale $L(p) \frac{\sqrt{p}}{M_0}$.
The general form of the contribution to real mass is  expected to be
$\Delta M(Coul)\sim \alpha_{em}\frac{M_0}{\sqrt{p(L)}}$. This motivates the
assumption that the  general p-adic form of the   Coulombic correction is

\begin{eqnarray}
\Delta M^2(Coul)&=&k(Coul)\alpha_{em}p^2\equiv Kp^2
\end{eqnarray}

\noindent where the parameter $k(Coul)$ is same for all particles of same
charge.  For $W/e$ and $\tau/e$ mass ratio the sign of the effect is
correct.  This cannot be the whole explanation: the point is that the
fraction of
 Coulombic energy from total energy is smaller for muon than for electron
and therefore the muon/electron mass ratio tends to decrease rather than
inrease.\\
 b) Second mechanism is related to the mixing of boundary topologies,
which turns out to have important role in quark physics.   The mixing of
$g=0,1,2$ topologies can affect the masses in second order only so that
in p-adic formalism one must have $sin^2(\phi_i)=n_i^2p$ for mixing
angles: this is possible  since $sin(\phi)\propto \sqrt{p}$ is possible if
4-dimensional algebraic extension is used.  The most general   mixing is
of form familiar from Kobayashi-Maskawa matrix

\vl
$\left[
\begin{array}{ccc}
c_1&s_1c_3 &s_1s_3\\
-s_1c_2& c_1c_2c_3-s_2s_3exp(i\delta)& c_1c_2s_3-s_2c_3exp(i\delta)\\
-s_1s_2& c_1s_2c_3+c_2s_3exp(i\delta) &c_1s_2s_3-c_2c_3 exp(i\delta)\\
\end{array}
\right]$

\vl

\noindent
 The sines $s_i$ of various angles are of form $n_i\sqrt{p}$ and
 phase phase factor is assumed to be $\epsilon=\pm 1$.   In extremely good
approximation the changes in p-adic mass squared  are given by

\begin{eqnarray}
\Delta X (e) &=& (K+9k_1 )p^2\nonumber\\
\Delta X(\mu)&=& (K-9k_1+51k_2 )p^2\nonumber\\
\Delta X(\tau)&=& (K+ 60k_2-9k_3)p^2\nonumber\\
k_1&=& n_1^2\nonumber\\
k_2&=& (n_2-\epsilon n_3)^2\nonumber\\
k_3&=& (n_2+\epsilon n_3)^2\nonumber\\
\end{eqnarray}
 \\
c) For electron the the contributions coming from Coulomb energy and mixing
 are compensated by the  change in the mass scale $m_0^2$.

\begin{eqnarray}
(m_0^2)_R &\rightarrow& \frac {5+{2}{3}}{5+\frac{2}{3}+\Delta (e)}\nonumber\\
\Delta  X(e) &=& ((K+9k_1)p^2)_R\nonumber\\
\
\end{eqnarray}

\noindent The latter formula holds true if $K+k_1$ is finite sum of powers
of two.\\ d) If one assumes that also the contributions to muon and tau
masses come  in powers of two
 one can write for the real counterparts of the masses the following
expressions

\begin{eqnarray}
\frac{m_2(L,exp) }{m^2(L,0)}&=&
\frac {s(L) +X(L) + \Delta (L) }{s(L) +X(L) } \frac{5+\frac{2}{3}}
{5+\frac{2}{3}+\Delta X(e)}
\nonumber\\
\
\end{eqnarray}

\noindent where $m^2(L,0)$ is the prediction for lepton mass without any
corrections. \\ e) Additional information to fix the values of the
parameters is needed  and one one might think that the requirement that
W/e mass ratio is predicted correctly gives   one constraint  but  this is
not all that is needed.  The constraint comes from the following
observation.
  It is natural to assume  that hadronic isospin splittings are of second
order in $p$ so that $s(H)$ is  same for all hadrons inside isospin
multiplets. If one fits the hadron masses using general formula provided
by Super Virasoro  representations one finds  that this requirement is
satisfied if the condition

\begin{eqnarray}
\Delta X(e)&\ge& 0.0155
\end{eqnarray}

\noindent is satisfied. The nearest power of two is

\begin{eqnarray}
\Delta X(e)&=& 2^{-6} \simeq 0.0156
\end{eqnarray}

\noindent  and leads to an excellent prediction for $Z$ boson mass

\begin{eqnarray}
m(Z,pred )&\rightarrow &\frac{5+\frac{2}{3}}
 {5+\frac{2}{3}+\Delta (e)} m(Z,0) \simeq .9986 m(Z,exp)\nonumber\\
\end{eqnarray}

\noindent  with error  of $0.24$ per cent and in the absence of other
corrections to $Z^0$ mass
 this means that the identification of mass scale must be correct. \\ f)
$W$ mass is very sensitive to the exact value of the Coulombic correction
$K$ and the requirement that $W$ mass is correct fixes the value of $K$

\begin{eqnarray}
m(W)& =& \sqrt{1+2K} \frac{5+\frac{2}{3}}
 {5+\frac{2}{3}+\Delta X(e)} m(W,0)\nonumber\\
K&=&2^{-4}(  1+2^{-2}+ 2^{-5} + 2^{-7} + 2^{-8}   ) \simeq 0.0808
\end{eqnarray}

\noindent   g)  Using the value $\Delta  X(e)$  and $K$ one can  deduce
the values of  the mixing parameters $k_1,k_2$ and $k_3$  from leptonic
masses and one obtains

\begin{eqnarray}
9k_1&=& \Delta X (e) - K =-(2^{-4}(  1+ 2^{-5} + 2^{-7} + 2^{-8}   )
\simeq -0.0652\nonumber\\
51k_2&=&   \Delta X(e) -2K +\Delta X(\mu)
= 2^{-7}(1+ 2^{-1}+ 2^{-2}+2^{-3}+2^{-5})  \simeq .0149\nonumber\\
9 k_3&=& \Delta  X (\tau) -\frac{60\Delta X(\mu)}{51} -
\frac{60\Delta X(e)}{51}
+ (\frac{120\Delta X (e)}{51}-1) K \nonumber\\
&=& -2^{-2}(1+ 2^{-1}+ 2^{-3}+ 2^{-6}+2^{-7}+2^{-8}
+ 2^{-9}+2^{-11}       ) \simeq -0.4293\nonumber\\
\
\end{eqnarray}

\noindent  The negative values of $k_1= n_1^2$  and $k_3$   do not mean
any  problem since $k_1$ is equivalent with $p-k_1$, which is positive
quantity. The real counterpart of
 $s_1^2$ can be defined  as  $(s_1^2)_R/((s_1^2)_R+(c_1^2)_R) $ and
corresponds to a rather
 large angle.

\subsection{Neutrino masses }

The estimation of neutrino masses is difficult at this stage since  the
prediction of the  primary  condensation level is not  yet possible.  The
cosmological bounds for neutrino masses however help to put upper bounds on
the masses. If one takes seriously the premilinary
data on neutrino mass measurement \cite{Israel} and the explanation of the
atmospheric  $\nu$-deficit in terms of $\nu_{\mu}-\nu_{\tau}$ mixing
\cite{Kamiokande,Russmix} one can
deduce that  the condensation level of all neutrinos is $k=163$ and deduce
information about the CKM matrix associated with neutrinos so that mass
predictions become  unique.

\vm

For neutrinos the expression for the  cm contribution is given by the
formula

\begin{eqnarray}
M^2(cm) &=&s \frac{M^2_0}{p}+X\nonumber\\
s&=&3 \nonumber\\
X&=&(-\frac{1}{2}p^2)_R
\end{eqnarray}

\noindent  where $M^2_0$ is universal mass  scale.   Second order
contribution depends on the value of $p$ since the p-adic inverse of the
number $20$ is contained in the formula (sub-index $R$ denotes real
couterpart of p-adic number).   Under rather general assumptions about
$p(\nu)$ the contribution can be written as

\begin{eqnarray}
M^2(int,\nu) &=&(3+  \frac{1}{2}) \frac{M^2_0}{p(\nu)}
\end{eqnarray}

 \noindent With the above described assumptions one has the
 following mass formula for neutrinos

\begin{eqnarray}
M^2(\nu)&=& A(\nu)(\frac{ M_0^2}{p(\nu)}\nonumber\\
A(\nu_e) &=& 3+ \frac{1}{2}\nonumber\\
A(\nu_{\mu})&=& 12+ \frac{1}{2}\nonumber\\
A(\nu_{\tau})&=& 63+ \frac{1}{2}
\end{eqnarray}

\noindent In \cite{padTGD} it was suggested that the primary  condensation
levels of neutrinos correspond to primes near prime powers  of $2$:
$p\simeq 2^k$, $k$ prime.  Together with known bounds for neutrino masses
this leaves only few possibilities for the values of  neutrino masses.

\vm

   Arguments based on 2-adic description of Higgs  mechanism (,which do
not hold true in present scenario)  suggest that the most probable condensation
levels
 correspond
to the primes

\begin{eqnarray}
k(\nu_e)&=&163\nonumber\\
k(\nu_{\mu})&=&  149\nonumber\\
k(\nu_{\tau})&=&137
\end{eqnarray}

\noindent    With this assumption

\begin{eqnarray}
m(\nu_e,163)&\simeq& 1.485  \ eV\nonumber\\
m(\nu_{\mu},149)&\simeq &0.305\ keV\nonumber\\
m(\nu_{\tau},137)&\simeq& 0.053 \ MeV
\end{eqnarray}

\noindent   These  estimates are
 consistent with the  recent  upper bounds  \cite{Bound} of order $10 \ eV$,
$270 \ keV$ and   $0.3 \ MeV$ for $\nu_e$,
 $\nu_{\mu}$ and $\nu_{\tau}$ respectively.

\vm

 The recent
measurement  \cite{Israel} suggests that the masses of both electron
and muon  neutrinos
are in the range $.5-5 \ eV$ and
that  mass squared difference is  $\Delta m^2=
m^2(\nu_{\mu})-m^2(\nu_e)$ is between $.25-25 \ eV^2$.  This requires
that $\nu_{\mu}$ and
 $\nu_e$
have common condensation level $k=163$ (in analogy with d and s quarks). The
resulting prediction (in absence of mixing) for $\nu_{\mu}$ mass is
$m(\nu_{\mu})=2.38 \ eV$ and  $\Delta m^2\sim 3 \ eV^2$.
 The interpretation of the experiment is based on
small  CKM  mixing of muon and electron neutrinos   having nearly degenerate
masses.
\vm

 The second source of information is the  atmospheric $\nu_{\mu}/\nu_e$ ratio,
which is roughly by a factor $2$ smaller than predicted by standard model
\cite{Kamiokande}. A possible explanation is the CKM mixing of muon neutrino
 with
$\tau$-neutrino, where as the mixing with electron neutrino is excluded as an
explanation. The latest results from Kamiokande \cite{Kamiokande} are in
 acordance
with the mixing $m^2(\nu_{\tau})-m^2(\nu_{\mu})Ê\simeq 1.6 \ 10^{-2} \ eV^2$
and mixing
angle $sin^2(2\theta)=1.0$: also the zenith angle dependence of the ratio
 is in
accordance with the mixing interpretation.  If this result is taken seriously
then   the only conclusion is that all three neutrinos condense on $k=163$
level
with $L(163) \simeq 7.6 E-7 \ m$.    Combining the
result of \cite{Israel} with the upper bound
$sin^2(2\phi)<.003$  for $e-\mu$ mixing angle  at large $\Delta m^2$ limit
 \cite{Russmix} one can conclude  that $e-\mu$ mixing is
small  as compared with $\mu-\tau$ mixing.

\vm

  From the
general form of the mass formula at $k=163$ level
and from the formula for the mixed masses

\begin{eqnarray}
M(\nu_{\mu})_{mix}&=& cos^2(\theta)
M^2(\nu_{\mu})+sin^2(\theta)M^2(\nu_{\tau})\nonumber\\
M^2(\nu_{\tau})_{mix}&=&
sin^2(\theta)M^2(\nu_{\mu})+cos^2(\theta)M^2(\nu_{\tau})\nonumber\\
\end{eqnarray}

\noindent one can deduce the value of mixing angle $\theta$
assuming $\delta M^2 (\tau-\mu)
=1.6 E-2 \ eV^2$ to be given. The predictions  are

\begin{eqnarray}
m(\nu_e)&=& 1.485 \ eV \nonumber\\
m(\nu_{\mu})_{mix}&=& 4.895 \ eV \nonumber\\
m(\nu_{\tau})_{mix}&=&4.891 \ eV\nonumber\\
sin^2(2\theta)&=& .99999949\nonumber\\
\Delta m^2(\mu-e)&=& 21.7 \ eV^2
\end{eqnarray}

\noindent The prediction of the mixing angle is in agreement with the
experimental results \cite{Kamiokande}. The prediction of mass for muon type
neutrino is  slightly below
the upper bound $5 \ eV$ given by  the previously mentioned experiment
\cite{Israel}.
Also the prediction for $\Delta m^2(e-\mu)$ is  within
preliminary  experimental bounds \cite{Israel}.
 A  more general  neutrino mixing matrix $ D$  of form

\vl

\begin{tabular}{||c|c|c|c||}\hline \cline{1-4}
&e &$\mu$ &$\tau$\\ \hline
$\nu_e$& $c_1$& $s_1$ &$0$\\ \hline
$\nu_{\mu}$& $-s_1c_2$ &$c_1c_2$ &$c_2$\\ \hline
$\nu_{\tau}$& $-s_1c_2$& $c_1c_2$ &$-c_2$\\
\hline\hline
\end{tabular}

\vl

\noindent  where $c_2=1/\sqrt{2}$ (masses of $\mu$ and $\tau$ neutrinos
are assumed to be identical) makes $\nu_e-\nu_{\mu}$ mass splitting smaller.
 For
$s_1^2=1$ one has
 $\Delta m^2(e-\mu)= 2.5\cdot 6 \ eV^2$.  Large value of $s_1$
 is questionable since it
is not at all obvious whether charged lepton mixings can be such that
 $\nu_{\mu}-e$
element of CKM matrix is of order $10^{-3}$  as required by \cite{Russmix}.

\subsection{Color excited leptons}

In the previous paper it was found that TGD predicts also color excited
leptons
 and quarks. Both charged and neutrinos allow massless $10$ and $\bar{10}$
color multiplets and neutrinos also $27$ dimensional color multiplet.
Also  $U$ type  quarks allow excitations created by  operators belonging
to  $10$ and $\bar{10}$  color multiplets.   In previous papers  the
existence of color octet (rather than decuplet) excitations of  leptons
were used to explain the anomalous $e^+e^-$ pairs observed in the heavy
ion collisions \cite{Heavy,Lepto}: the resonances were identified as color
bound states of colored leptons. Effectively the hypothesis
 means  the existence of a new branch of Physics at the energy scale of
one MeV.  The
 decay widths of the intermediate gauge bosons seem to exclude light
exotic fermions  and in the fourth paper of the series it will be shown
how p-adic unitarity allows to  avoid this restriction by replacing the
condition on the number of light fermions with
 modulo type condition.

\section{Masses of elementary  bosons}

The explicit calculations show that the masses of gauge bosons are in
 qualitative accordance with expectations assuming $T(ew)=1/2$  for
electroweak  gauge bosons and $T=1$ for exotic bosons. \\ a)  Gluons are
predicted to be exactly massles to order $O(p^2)$ whereas
 photon has extremely small thermal mass. The requirement that photon is
massless fixes the parameter $k(B)$ to $k(B)=3/2$, which is the ratio of
Ramond and ground state N-S mass scales. The difference between N-S and
Ramond string tensions  implies that ground state mass scales  are
identical for these representations. \\   b)   Weinberg angle is predicted
to be $sin^2(\theta_W)= 3/8$, typical value for the parameter in GUTs at
symmetry limit.  Secondary topological condensations are expected to
renormalize the value of the Weinberg angle in TGD.  \\ c)   The
prediction for the ratio of $W$ boson and electron masses is too small by
$-0.7$ per cent.  Main discrepancy is related to the too high value of
$Z^0$ mass implied by the too large value of Weinberg angle.  It was
already found that these discrepancies are disappear,   when topological
mixing effects for leptons and Coulomb energy are taken into account.
 The requirement that $sin(\theta_W)$ and $cos(\theta_W)$ are rational
numbers is very attractive and implies that Weinberg  angle corresponds to
Pythagorean triangle. Requiring that the sides of this triangle are as
small as possible one obtains the  triangle $(r,s)=(4,1)$ and  $P=
8^2/17^2 \simeq 0.2215$.  For this value of Weinberg angle $Z^0$ mass is
predicted within one per cent error correctly.  The Pythagorean value of
Weinberg angle is within experimental uncertainties identical with the
value of Weinberg angle $ P= 0.2218\pm 0.0059 $ determined from
neutrino-nucleon scattering \cite{Nenucl} and slightly smaller
 than the value $0.2255\pm 0.0019$ determined from LEP precision
experiments  \cite{Fortschritte}.

\vl

\begin{tabular}{||c|c|c|c||}\hline\hline
boson&$M_{obs}/MeV$& $M_{pred}/MeV$&error/\%\\ \hline \hline
$\gamma$&0&0&0\\ \hline\hline
gluon&0&0&0 \\ \hline\hline
 $W$ &80200&79582&-0.8\\ \hline\hline
$Z$ &91151&100664&10\\ \hline\hline
\end{tabular}

\vl

Table 2.2. \label{Bosonmasses}  Masses of nonexotic gauge bosons without
corrections coming from topological mixing of leptons and Coulomb energy.

\vl

\noindent  d) There is large number of candidates for exotic bosons.   If
one assumes $T=1$ for exotic bosons most
  states  possess either Planck mass or are massive but 'light'.   What is
remarkable is the absence of massless noncolored exotic bosons for $T=1$.

 \vl

\begin{tabular}{||c|c|c|c||}
 \hline\hline
spin  &charge operator &$ M^2 (T=1)$ &$ M^2 (T=1/2)$  \\ \cline{1-4}\hline
0&$I^3_{L/R}$& Planck mass&$\frac{1}{2p}$\\ \hline
0&$I^{\pm}$ & Planck mass &$\frac{1}{2p}$  \\ \hline
0&$I^3_{L/R}Q_K$ &$\frac{3}{p}$&0\\ \hline
0&$I^{\pm}Q_K$ &Planck mass&$\frac{1}{2p}$\\ \hline
1&$1$ &Planck mass&$\frac{1}{2p}$\\ \hline\hline
\end{tabular}

\vl

Table 2.3. \label{Exobosonmasses} Masses and couplings of noncolored light
 exotic
bosons for $T=1$  and $T=1/2$.  Charge operator
 tells how the boson in question couples to matter. For $T=1/2$
the states with charge operator $I^3_{L/R}Q_K$ are essentially
massless for large values of $p$ and some additional light states
 become possible.

\vl

\begin{tabular}{||c|c|c|c|c||}
 \hline\hline
spin  &charge operator & $D$ &$ M^2 (T=1)$&$M^2(T=1/2)$ \\ \cline{1-5}\hline
0& $I^{\pm}Q_K$ &8 &$\frac{2}{p}$&0\\ \hline
1& $I^{\pm}$ &8 &Planck mass&$\frac{1}{2p}$\\ \hline
1& $I^3_{L/R}Q_K$ &8 &Planck mass&$\frac{1}{2p}$\\ \hline
1&$I^{\pm}Q_K$&8 &Planck mass &$(\frac{3p}{10})_R\frac{1}{p}$
\\ \hline\hline
0& $I^{\pm}Q_K$ &$10,\bar{10}$ &$\frac{3}{p}$&0\\ \hline
0& $1$ &$10,\bar{10}$ &0&$0$\\ \hline
1&$I^{\pm}$&$10,\bar{10}$ &$\frac{3}{p}$&0\\ \hline
1&$I^3_{R/L}Q_K$&$10,\bar{10}$ &$\frac{3}{p}$&0\\ \hline\hline
0&$I^{\pm}$&27 &Planck mass &$\frac{1}{2p}$\\ \hline
0&$I^3_{R/L}Q_K$&27 &Planck mass &$\frac{1}{2p}$\\ \hline
1&$I^3_{R/L}$&27 &Planck mass &$\frac{1}{2p}$\\ \hline
1&$Q_K$&27 &Planck mass &$\frac{1}{2p}$\\ \hline\hline
0&$I^3_{R/L}$&27 &0 &0\\ \hline
0&$Q_K$&27 &0 &0\\ \hline
1&$1$&27 &0 &0\\ \hline\hline
\end{tabular}

\vl

Table 2.4. \label{Colbosonmasses} Masses and couplings of colored
 exotic
bosons for $T=1$ and $T=1/2$.  The last two massless bosons are doubly
 degenerate due to occurrence of two $27$-plets with conformal weight
$n=2$. $T=1/2$  is physically possible alternative since no long range forces
are implied.

\vm

\noindent e) There is distinct possibility for the higher generation gauge
bosons. The higher generations of intermediate gauge bosons should have
condensation level with $p<M_{89}$:  the first candidate is $p\simeq
2^{83}$: the masses of $g=1$ bosons are for $k=83$   $m(W(g=1))\simeq 24.0
m(W)$ and $m(Z(g=1))\simeq 19.6 m(Z)$.  The next possibilities are
$k=79,73,71$ and $M_{61}$.  Also photon could allow higher generation
excitations: they would be certainly massive and have mass
$\sqrt{2N(g)g/p(\gamma)}$. \\ f) There is no boson identifiable as Higgs
doublet:  doublet is  excluded already by the fact that $N-S$
representations do not allow isospin doublets.  The result is  in
accordance with
 earlier suggestions.  The TGD:eish counterpart of Higgs field is
Virasoro  generator  $L_0$, which develops vacuum expectation value in
conformal symmetry breaking description of Higgs mechanism.\\  d)
Graviton is not possible  in the proposed scenario. The largest possible
values of spin and isospin are $J=1$ and $I=1$.  This result is in
accordance with the basic assumptions about  p-adic TGD as flat spacetime
limit of Quantum TGD.  Of course, one could try to construct the
counterpart of closed string model (closed string is replaced by two
sheets of flat spacetime glued along boundaries) to get graviton but there
is no guarantee that  spin $2$ massless state having $D(1)=0$ is obtained.

\section{Hadron  masses}

The general mass formula can be written as  sum of interior and boundary
contributions:

\begin{eqnarray}
M^2(h) &=&  M^2(\delta)+  M^2(int) \nonumber\\
M^2(\delta)&=& M^2(cm)+M^2(mod)
\end{eqnarray}

\noindent  The contribution $ M^2(mod)$ from  the modular degrees of
freedom is
 same as for leptons and sum over the contributions of hadronic valence
quarks.   Boundary cm contribution can be evaluated using p-adic
thermodynamics and it turns out that the contribution is in first order
just the sum of single quark contributions and of same form as for
leptons.  In second order mass squared  is   not any more additive. For D
quark second order contribution is same as for charged leptons but for U
quark the contribution is not identical with that of neutrino.

\vm

The mere boundary contribution gives mass formula, which is  qualitatively
correct but  it is obvious that something is missing.  There are several
mixing effects present.\\ a) It turns out that modular  contribution
dominates for heavier quarks and predicts too large masses for hadrons
unless  mixing of boundary topologies $U$ and $D$ type quarks is allowed.
Different topological mixing for $U$ and $D$ type quarks implies  in turn
nontriviality of Kobayashi-Maskawa matrix.  Although one cannot predict at
this stage the mixing angles one can use some empirical input together
with number theoretical consideration to derive strong constraints on the
values for mixing angles.  First, the change  $\Delta s$ of quark is small
integer. p-Adic unitarity plus the assumption that $\vert U_{ij}\vert^2$
are rational numbers (allowing  interpretation either as real or  p-adic
numbers)  implies that the mixing matrix is determined by two small
parameters.  More stringent condition is rationality of $U_{ij}$.  Cabibbo
angle is predicted correctly within experimental uncertainties from the
requirement that $u$ and $d$ quarks have identical masses in order $O(p)$
and the orders of magnitude for the small elements of the KM matrix are
predicted correctly.    Even the necessity of nontrivial phase factors
leading to CP breaking is forced by number theory.  It should be noticed
that in TGD framework  the mixings of $U$ and $D$ type quarks are both
observable, not only the difference between these mixings as in standard
model. \\ b)  Mixing  of boundary topologies is not enough: the masses of
hadrons
 containing one charmed or bottom quark are systematically too heavy
whereas masses of diagonal mesons of type $c\bar{c}$ and $b\bar{b}$ are
predicted quite satisfactorily.   The explanation is mixing of primary
condensation level for $c$ and $b$ quark. $c$ and $b$ do not spend all
their time at $k=103$ condensation level but condenses now and then on
$u,d,s$ quark having $k=107$. Condensation is not possible in diagonal
mesons since no $u,d$ or $s$ type quark is present. \\ c)  Third type of
mixing effect is mixing of neutral pseudoscalar mesons made possible by
annihilation to two-gluon intermediate state.  The mixing is important for
mesons such as $\eta,\eta^,\eta_c$ and in the  absence of  the mixing
$\eta$ and $\pi$ would have identical masses.

\vm

The remaining  hitherto  identified $O(p)$ contributions come from the
interior of hadron.  \\ a)  Color Coulombic force is strong and is
 expected to give sizable contributions to mass squared.  Since $T=1$
turns out to be the only sensible choice for the p-adic temperature the
value of $s(M)$ should be nonvanishing for all mesons.  For pion the mass
fit however gives $s(\pi)=0$.  This means that interior contribution to
mass squared of mesons  must be  of the order $O(p)$, be negative and
integer value  and cancel the contribution  coming from quark masses in
pion.  Contribution is also same for all baryons/mesons.
   \\ b)  Color magnetic hyper fine splitting  gives nice explanation for
$\rho-\pi$,  $K^*-K$, $ N-\Delta$ mass splittings and must be included
also now in order $O(p)$.   Number theoretic
 considerations force this contribution to be  sum of integer valued
contributions. The integers
 associated with each quark pair are in principle different (contribution
is inversely proportional  to the product of quark masses in first order
QCD).    \\ c)  It turns out necessary to assume
 isospin-isospin interaction between baryonic quarks of same generation in
order to understand masses of  doubly strange baryons. The general form of
the splitting is same as for color magnetic interaction.
 For mesons this interaction is in principle also present but turns out to
vanish in first order.  Also isospin isospin interaction between different
generations is in principle present but can be
 assumed to vanish.

\vm

Second order contributions to hadron masses are small as compared to
$O(p)$  contributions with pion forming exception:  the prediction of pion
mass gives a  crucial test for the model in second order.
 Perhaps the most important second order contribution comes from boundary
cm  degrees of freedom and this contribution is isospin dependent and
depends on the electroweak isospin of the state only.  This contribution
is identical for neutral and charged pion and about 3 per cent smaller
than pion mass.  The inspection of hadronic isospin splittings shows that
this contribution cannot be the only one.  The   ordinary electromagnetic
Coulomb interaction is expected to give an additional negative second
order contributions.

\vm

The contribution of sea partons is expected to be present, too.  The case
of  pion suggests that contribution is small as compared to second order
contribution given by quark masses.  An attractive possibility is  that
thermodynamic description  applies to sea partons, too.   In thermodynamic
description one must  introduce chemical potential for each parton type
and in low temperature limit $\mu/T$ is quantized to integers  by p-adic
existence requirement for Boltzmann weight: $\mu/T=n$.  This fixes
entirely the number distribution for sea apart from the parton
distribution functions for longitudinal momentum fraction (recall that
quarks are most of the time massless particles!).   For nonvanishing value
of chemical potential sea partons give only $O(p^2)$ contribution to mass
squared.

\vm

It must be  admitted that TGD doesn't yet predict the various parameters
  needed for understanding hadronic masses. It must be emphasized however
that by number theory all  these parameters are  just integers with very
limited range of allowed values and simple physical constraints turn out
to make the choice between the alternatives unique.  The errors of
resulting simple mass formula are  about one or two  per cent.

\subsection{Evaluation of boundary cm contribution to hadron mass}

The contribution coming from the cm of boundary component  can be
evaluated  using p-adic thermodynamics for the tensor product of quark
Super Virasoro representations.  Since quarks correspond to different
boundary components and Super Virasoro generators correspond to
infinitesimal conformal  transformations acting in each boundary component
separately it is obvious that quarks must obey Super Virasoro conditions
separately.   Since the value of first order contribution to mass squared
is  typically of the order of Planck mass it would be highly desirable if
quarks would obey same mass formulas as leptons in first order.  The
necessary condition for  this  is that   isospin  splitting of quark mass
squared is described as the shift of vacuum weight $h$
 using same formula for leptons and quarks

\begin{eqnarray}
h(U)&=& -2 -2/9+\frac{Q_K^2}{2}=  -2\nonumber\\
h(D)&=&-1
\end{eqnarray}

\noindent  so that essentially same operators create $M^2=3/2$ states for
$U$
 and $\nu$ and $D$ and $e$ respectively.

\vm

The general p-adic  mass formula was derived from p-adic thermodynamics in
 previous paper and is given by

\begin{eqnarray}
\frac{M^2(h)}{M_0^2}&=& sp + Xp^2\nonumber\\
s&=& k(F)\frac{D(h,1}{D(h,0)}\nonumber\\
X&=&k(F)(2D(h,2)-(\frac{D(h,1)}{D(h,0)})^2) \nonumber\\
k(F)&=&3/2
\end{eqnarray}

\noindent Here $D(h,n)$  denotes  the degeneracy of state with mass
$M^2=3/2n$
 created by gauge invariant operators of conformal weight $n$ from vacuum.

\vm

For hadron state with mass $M^2=0$ has degeneracy given by the product of
single
 quark degeneracies $D(h,0)$

\begin{eqnarray}
D(h,0)&=& \prod D(q_i,0)
\end{eqnarray}

\noindent  $M^2=3/2$ states of hadron are  obtained by exciting one of the
quarks to this state while other quarks remain massless.  This gives

\begin{eqnarray}
s(h) &=& \sum_i \frac{D(1,i)}{D(0,i)}= \sum_i s(q_i)
\end{eqnarray}

\noindent so that mass squared is additive in first order. This result
implies that hadrons are light.
 The allowance of color excitations for individual quarks would change
the degeneracies of both $M^2=0$ and $M^2=1$ states and as a result most
hadrons would have Planck mass.

\vm

The values of the parameters $ s(q)$

\begin{eqnarray}
s(U)&=&s(\nu)=3\nonumber\\
s(D) &=&s(e)=5
\end{eqnarray}

\noindent  were derived in previous paper and are same as for leptons.  The
 first formula implies  large isospin splitting for $u$ and $d$ quark and
the mixing of boundary topologies for $U$ and $D$ quarks must be such that
effective values of $s(U)$ and $s(D)$ are identical.  The requirement
implies essentially the experimental value for Cabibbo angle.

\vm

$M^3=3$ excitations correspond to states were single quark is excited  to
$M^2=3$ state or   to states were two quarks are excited to $M^2=3/2$
state  both.   This means that mass squared is not additive in second
order. The general  expression for $D(h,2)$ is given by

\begin{eqnarray}
D(h,2)&=& \sum_i D(2,q_i) + \sum_{pairs \ (i,j)} D(q_i,1)D(q_j,1)
\end{eqnarray}

\noindent and one obtains for the coefficient of the second order
contribution the following expression

\begin{eqnarray}
X(h) &=& \sum _i X(q_i)+  \Delta (h)\nonumber\\
 X(q_i)&=&   \frac{D(q_i,2)}{D(q_i,0)} - \sum_i(\frac{D(q_i,1)}
{D(q_i,0)})^2
\nonumber\\
\Delta  (h) &=&  \sum_{pairs}  \sum \frac{D(q_i,2)}{D(q_i,0)} +
 \sum_{pairs} \frac{D(q_i,1)D(q(_j,1)}{D(q_i,D(q_j,0)} -(\sum
\frac{D(q_i,1)}{D(q_i,0)})^2 +  \sum_i(\frac{D(q_i,1)}{D(q_i,0)})^2
\nonumber\\
\
\end{eqnarray}

\noindent   The term $\Delta$ gives deviation from the simple additive
 formula for mass squared.  The values of the coefficients $D(U,n)$ and
$D(D,n)$ were derived in  previous paper and are given by the table below.

\vl

\begin{tabular}{||c|c|c|c||}\hline\hline
quark type &$ D(0)$& $D(1)$&$D(2)$\\ \hline \hline
U&40&80&8\\ \hline\hline
D&12&40&80 \\ \hline\hline
 \end{tabular}

\vl

Table 4.1.  \label{Quarkdege} Degeneracies for various mass squared values
for
 quarks.

\vm

As far as cm contribution to mass is considered there are only 4
nonequivalent
 combinations  for baryonic quarks, namely  $UUU$,$UUD$,$UDD$ and $DDD $.
  For mesons there are 3 combinations $ U\bar{U}$, $D\bar{D}$ and
$U\bar{D}$.  The integer
 part  of $X$ gives completely negligible contribution to mass squared in
the absence of
 mixing of different quark configurations ($D\bar{D}$ and $U\bar{U}$ for
neutral pion).
 The following table summarizes the values of  first and second order
terms $X$
 for these configurations. Also are listed the real counterparts for
$Xp^2$  obtained by using the canonical correspondence between p-adic and
real numbers ($\sum x_kp^k\rightarrow \sum x_kp^{-k}$).

\vl

\begin{tabular}{||c|c|c|c||}\hline\hline
hadron \ type&s&X& $p(Xp^2)_R$\\ \hline \hline
UUU&9&  $\frac{24}{30}$&   $\frac{24}{60}$             \\ \hline\hline
UUD&12&    $\frac{13}{30 }$&  $\frac{13}{60} $        \\ \hline\hline
 UDD &13&   $ \frac{18}{30}$  &  $\frac{8}{60}$     \\ \hline\hline
DDD &15&      0  &      0   \\ \hline\hline
$U\bar{U}$&6& $\frac{6}{30}$&  $ \frac{4}{5}$    \\ \hline\hline
$D\bar{D}$&10&  $-\frac{10}{30 }$  & $\frac{1}{3}$\\
\hline\hline
$U\bar{D}$&8&   $\frac{13}{30}$  & $\frac{2}{15}$    \\ \hline\hline
\end{tabular}

\vl

Table 4.2.  \label{HadroX} The values of the second order contribution to
mass  squared coming from quark masses for various quark combinatios.

\vm

 The mass fit for pion shows that $s(eff)$ must vanish for pion  (due to
interior contributions to mass squared) so that second order
 contribution should give entire pion mass.   Neutral pion is
superposition  of form $\sqrt{1/2}(U\bar{U}+D\bar{D})$  one  obtains for
the masses of neutral and
 charged pions the following expressions

\begin{eqnarray}
X(\pi^0)&=& \frac{1}{2}(  X(U\bar{U}) +X(D\bar{D})=(\frac{3}{5}-
\frac{1}{6})
 =\frac{13}{30}\nonumber\\
X(\pi^+)= X(U\bar{D})=\frac{13}{30}
\end{eqnarray}

\noindent It should be noticed that   in the calculation of pion mass one
cannot
 use directly the sum of  parameters $X$ associated with $U\bar{U}$ and
$\bar{D}{D}$ since this would drop  the possible half odd integer
contribution to $X_{eff}$.   It is remarkable that the predicted values
are identical for neutral and charged pion.
 In order to calculate the real mass one must find the real counterpart
of $Xp^2= (13p^2)/30$ by canonical correspondence between p-adic and real
numbers.  Using the formulas from the last appendix  of previous paper one
has
 $ Xp^2\rightarrow X_R= 22/60p$ for $p=M_{107}$.  The result of mass fit
is $X_R= 24/p$ for charged and $X_R=23/p$ for neutral pion. Errors are of
order 3 per cent!

\subsection{Contribution of modular degrees of freedom and primary
condensation levels}

 Modular degrees of freedom give readily calculable real  contribution
$M^2(mod)= 3gN(g)M_0^2/p(q)$ to the mass. $p(q)$ is the prime associated
with the primary condensation level of the quark and depends on quark.
Since tensor product of Super Virasoro representations associated with
 boundary degrees of freedom is involved in calculation the
  boundary contributions of quarks to total mass  squared are  summed
together already  at p-adic level. The only possibility is that $k=107$ is
common    secondary   condensation level of $M_{107}$ hadron
 physics whereas
 primary condensation levels must be differ from $k=107$ for heavier
quarks.  The only possible primary condensation levels  for quarks turn
out to be following  ($p(q) \simeq 2^k$):

\begin{eqnarray}
k(u)=k(d)=k(s)&=&107\nonumber\\
k(c)=k(b)&=&103\nonumber\\
k(t)&=&89 \ or \ 97
\end{eqnarray}

\noindent The secondary and primary condensation levels of $u,d,s$ quarks
are identical.  This sounds peculiar at first but is in accordance with
the mass minimization mechanism suggested in second paper of series: if
primary and  secondary condensation levels have nearly identical
 $p\simeq 2^k$ then second order contribution to mass can suffer large
decrease in secondary condensation.
 In \cite{padTGD} the existence of an entire hierarchy of hadronic physics
 was suggested in the sense that there is hadronic physics associated with
each Mersenne prime.  In particular,  quarks of each generation have
replicas with scaled up masses.

\subsection{Mixing of boundary topologies}

In TGD the different mixings of boundaries  topologies for $U$ and $D$ type
quarks provide the fundamental mechanism for Cabibbo mixing and also CP
breaking.  In the determination of Kobayashi-Maskawa matrix one can use
following  conditions. \\ a) Mass squared expectation values  in order
$O(p)$  for mixed states
 must be integers and the study of hadron mass spectrum leads to very
stringent conditions on the values of these integers. Physical values for
these integers imply essentially correct value for Cabibbo angle provided
$U$ and $D$ matrices differ only slightly from mixing matrices  mixing
only  the two lowest generations. \\ b)  The matrices $U$ and $D$
decribing the mixing of $U$ and $D$ type boundary topologies are unitary
in p-adic sense.  The requirement that the moduli squared  of the matrix
elements are rational numbers is very attractive since it suggests
equivalence of p-adic and real probability concepts  and therefore could
solve the conceptual problems related  to   the transition form p-adic to
real regime.  It must be however immediately added that rationality
assumption for  the probabilities defined by S-matrix turns out to be
unphysical.
 Unitarity requirement  is  nontrivial since mass conditions give
constraints  for the squares of cosines and sines for various angle
parameters, only. The requirement that sines and cosines  parametrizing U
and D matrices exist as p-adically real numbers is very strong and leads
to the conclusion that $U$ and $D$ matrices contain only two small
parameters.   \\ c) One can  require that the probabilities defined CKM
matrix elements are also  rational numbers or even that U,D and CKM
matrices are rational and this gives very strong number theoretic
conditions.  The general  solution for CKM  rationality  conditions can be
found.  Two of the angles
 appearing in U and D matrix correspond to Pythagorean triangles and the
remaining two angles to   triangles with integer valued shorter sides
with  sine and cosine  allowing common irreducible rational  phase.  If
the elements of U,D and CKM matrices are required to be rational all angle
parameters correspond to Pythagorean triangles. This alternative is not
excluded by number theoretic condition
 in  the scenario providing a solution for the spin crisis of proton.
 \\ d) The requirement that Cabibbo angle has correct value fixes
essentially uniquely the values of mixed mass squared for various  mesonic
quark  generations.  In real regime it is quite easy to reproduce
KM-matrices   satisfying experimental
 constraints by choosing appropriate values for the small parameters, one
of which is CP breaking angle but the problem whether there actually
exists any U and D  matrices satisfying number theoretical conditions
remains open at this stage.

\subsubsection{ Do mesonic and baryonic quarks mix identically?}

The attractive assumption that the  mixing matrices for baryons and
mesons  are identical  gives for  Cabibbo angle  the rough estimate
 $sin(\theta_c)\simeq 0.2236$, which is slightly below the experimentally
allowed range $sin(\theta_c)=  0.226\pm 0.002$: mixing with third
generation can slightly increase the value.   This potential
 discrepancy could serve as motivation for asking whether the mixing
matrices could be different for
 baryons and mesons (to be honest,  the actual motivation was
calculational error, which yielded quite too
 small value of Cabibbo angle!)

\vm

One could indeed seriously consider the possibility that topology mixing
and
 perhaps even CKM matrices
 are different in baryons and mesons.
 The  point is that it is the electroweak current,  which contains  the
Cabibbo  mixed  quarks.   The matrix elements of the  electroweak and color
currents are not changed between baryon states if one performs any unitary
change of basis.  In  GUTS the mixing of $U$ and $D$ type quarks
corresponds to global gauge transformation and therefore unitary
transformation.    In TGD the  mixing of quarks is not a gauge symmetry
since mass squared changes for quark but for all amplitudes expressible as
matrix elements of currents there are no observable effects.  The
different mixings for mesons and baryons are not seen at the level of
amplitudes by baryon number conservation at single particle level, say in
semileptonic decays of baryons.

\vm

The first instance, where effects might be seen is $B\bar{B}$ ahhihilation
to meson pair but if  the amplitude for this  process can be  described
in terms of electroweak and colored currents there are no observable
effects.  This is not obvious in TGD.  If dual diagrams provide  more
than  a phenomenological   description of nonperturbative aspects of this
process then $d$ and $s$ quark lines running from meson to baryon must
contain a vertex, where mixing angle and quark mass changes and empirical
effects are prediced.   Of course, the troublesome fact that mesonic and
baryonic quark masses tend to be different also in naive quark model  fit
of hadron  masses can be regarded as direct indication of correctness of
the TGD:eish prediction.       The appearence of  mesonic Cabibbo mixing
at the level  of currents follows from the fact   electroweak currents
have same electroweak quantum number structure as vector mesons.   For
example,  the electroweak couplings of the  charged electroweak current to
baryons  at low energies  can be in good approximation expressed using
generalized vector dominance model:  the emission of gauge boson from
baryon decomposes to the emission of vector meson, which couples to gauge
boson.  This means that matrix elements of currents are   proportional to
the matrix elements of meson fields.

\vm

The calculations show that effective mass squared for strange quark is
different  for mesons and baryons in the optimal scenario so that
baryonic and mesonic  mixing matrices
 are different.    CKM matrices  are  however  essentially identical for
the
 physical values of parameters.

\subsubsection{Topology mixing and quark masses}

The requirement that hadronic mass spectrum is physical requires  mixing
of $U$ and $D$ type boundary topologies. The following arguments fix the
effective values of the modular contribution $s(\delta,q_i)$  to the mass
squared.

\vm

\noindent  a) The smallness of isospin splitting for hadrons containing
only
 $u$ and $d$ quarks implies that the effective value of the modular
contribution  $s(\delta)$  must be same for $u$ and $d$ quark:
$s_{eff}(u)=s_{eff}(d)$ for
 all hadrons. \\ b)  The requirement that $\Delta -p$ mass difference
resulting from hyper fine splitting and isospin-isospin interaction is of
correct size  requires $s_{eff}(u)=s_{eff}(d)=s=8$ in baryons.  If color
binding energy is required to   be nonvanishing for baryons as it is in
mesons then
 $ s(d)=s(u)=9$ is the only possibility.   \\ c)  Proton-$\Lambda$  mass
difference is not affected by colormagnetic  spin-spin splitting nor
isospi-isospin interaction and suggests strongly
  $s_{eff}(s)=16$ in baryons. Also $\Omega$ mass suggests same.  For
mesons  $s_{eff} (s)\ge 13$ is forced by the need to obtain mixing
between  third and lower generations at all.
 $K-\pi$ mass difference suggests $s_{eff}(s)=14$.  This raises the
possibility  that  baryonic and mesonic  mixing matrices are different: it
turns out that for  physical parameter values  CKM matrix depends only
very weakly on
  $s_{eff}(s)$.   \\ d)    The masses  $c\bar{c}$ mesons come out
correctly if one has
 $s_{eff}(c)=6$. This alternative is also forced by the need to get mixing
of third generation with the lower ones.  For baryons same value works
best. \\ e)  Unitarity requirement fixes the values of $s_{eff}(b)$ and
$s_{eff}(t)$:  one must have $\sum_i s_{eff}(q_i)= 69$. \\ f)  The
assumption $s(u)=6$ looks at first rather crazy: if other angles are
small $u$ quark would spend approximately time fraction $3/9$more time
$g=1$ state than  in $g=0$ state!  This could
 however solve the spin crisis of the proton: the fraction of parton spin
from the
 spin seems to be only $30$ per cent. Assume that  the measured  non
strange parton spin corresponds  to $g=0$ spin fraction in proton.  If
$ud$ pair is in spin singlet state then  its contribution to spin vanishes
and only the
 contribution of $d$ remains, which gives approximate  spin fraction
$(9-s(u))/9=1/3$ for $s(u)=6$: agreement
 is quite good!   The corrections coming  from the mixing with higher
generation slightly  increase the fraction of $g=0$ spin.   For neutron
corresponding result is
 $(9-s(d))/9=5/9$ maximal mixing  scenario.

\vm

To summarize:\\
 The scenarios  $(n_1(d)=3, n_2(d)=8,n_1(u)=5, n_2(u)=6)$  and $(n_1(d)=4,
n_2(d)=8,n_1(u)=6, n_2(u)=6)$ favoured by spin crisis of proton  are the
best alternatives as far as masses are considered and  predict
 same Cabibbo angle at the limit,   when mixings with  the third
generation is  absent. CKM matrix turns out to depend only very weakly on
$n_{eff}(s)$
 and $K-\pi$ mass difference suggests that value $n_{eff}(s,M)=9<11$.

\vl

\begin{tabular}{||c|c|c|||}\hline\hline
 quark&$n$&$n(eff)$\\ \cline{1-3}\hline
d&0&(3) 4 \\ \hline
s&9&11\\ \hline
b&60&(55) 54 \\ \hline
u& 0&(5) 6\\ \hline
c& 9& 6\\ \hline
t&60&(58) 57 \\ \hline\hline
\end{tabular}

\vl

Table 4.3. \label{KMscenario}  The scenario $(n_1(d)=4, n_2(d)=11,n_1(u)=6,
 n_2(u)=6)$   is forced by
 sensible mass spectrum for baryons and and spin crisis of proton. The
values of parenthesis correspond to competing scenario, which doesn't
however solve  the spin crisis and is not consistent with rationality of
U and D matrices.

\vl

 Mass  constraints give  for the D matrix the following conditions

\begin{eqnarray}
 9\vert D_{12}\vert ^2 + 60 \vert D_{13}\vert ^2 &=& n_1(d)\nonumber\\
9\vert D_{22}\vert ^2 + 60 \vert D_{23}\vert ^2 &=& n_2(d)\nonumber\\
9\vert D_{32}\vert ^2 + 60 \vert D_{33}\vert ^2 &=&
n_3(d)=69-n_2(d)-n_1(d)\nonumber\\
\
\end{eqnarray}

\noindent The third condition is not independent since the sum of the
conditions is
 identically true by unitarity.

\vm

For $U$ matrix one has  similar conditions.
The task is to find unitary mixing matrices satisfying these conditions.

\subsubsection{The general form of U and D matrices}

The general form of U and D matrices is taken to be same as the standard
 parametrization of
Kobayashi-Maskava matrix.

\vl

$\left[
\begin{array}{ccc}
 c_1&s_1c_3 &s_1s_3\\
-s_1c_2& c_1c_2c_3-s_2s_3exp(i\delta)& c_1c_2s_3-s_2c_3exp(i\delta)\\
-s_1s_2& c_1s_2c_3+c_2s_3exp(i\delta) &c_1s_2s_3-c_2c_3 exp(i\delta)\\
\end{array}
\right]$

\vl

Table 7. \label{KMtype} CKM type parametrization for U and D matrices.

\vm

\noindent  Similar parametrization applies to $U$ matrix.   One can
multiply  the rows and columns of $U$ and $D$ with constant phases. Only
the multiplication of the columns of $U$ matrix affects KM matrix defined
as

\begin{eqnarray}
V&=& DU^{\dagger}
\end{eqnarray}

\noindent  It is natural to require that the squares of the moduli of
 $D_{ij}$ and $U_{ij}$ are rational numbers of equivalently:  the squares
of sines and cosines appearing in $D$ and $U$ are rational numbers.

\vm

The parametrization used guarantees only formally unitary. Mass conditions
 give constraints on the values of squares of cosines and sines and the
requirement that cosines/sines exist as p-adically real numbers is not
trivially satisfied.

\subsubsection{Explicit treatment of mass conditions}

The mass condition for d (u) quark gives following constraint between
 $s_1$ and $s_3$

\begin{eqnarray}
s_3^2&=&   \frac{(n_1-9s_1^2)}{51s_1^2}
\end{eqnarray}

\noindent where $n_1=3$ for $d$ and $n_1=5$ for $u$.

\vm

The mass condition for $s$ (c) quark gives following condition

\begin{eqnarray}
c_2^2X- V&=& -c_2s_2Y \nonumber\\
X&=& 51((c_1s_3)^2-c_3^2)-9s_1^2\nonumber\\
 Y&=& 102c_1c_3s_3c_{CP}\nonumber\\
V&=& n_2-9 -51c_3^2
\end{eqnarray}

\noindent One can reduce this equation to order equation for $c_2^2$ by
taking squares of both sides. The solution of the resulting equation reads
as

\begin{eqnarray}
c_2^2&=& -\frac{b}{2} +\epsilon_1 \sqrt{b^2-4c}\nonumber\\
b&=& -\frac{Y^2+2VX}{X^2+Y^2}\nonumber\\
c&=& \frac{V^2}{X^2+Y^2}\nonumber\\
\epsilon_1&=&\pm 1
\end{eqnarray}

\noindent p-Adicity gives nontrivial additional constraint. The argument
of square root must be square of a rational number

\begin{eqnarray}
b^2-4c&=& (m/n)^2
\end{eqnarray}

\noindent This condition in turn gives second order equation for $Y^2$,
 which can be solved and gives

\begin{eqnarray}
Y^2&=& -2V(X-V)+\epsilon_2 2V\sqrt{  (X-V)^2 +4(\frac{m_1}{n_1})^2}
\end{eqnarray}

\noindent Here $(m_1/n_1)$ is related to $m/n$ by $(m/n)= 2V(m_1/n_1)$.
Additional consistency condition results from the requirement that also
the argument of this square root is square of a rational number.  The
condition reads as

\begin{eqnarray}
(X-V)^2&=& (\frac{k}{l})^2 -( \frac{m_1}{n_1})^2
\end{eqnarray}

\noindent  The condition states that $X-V$ is a rational number, whose
square is difference of two squares of rational numbers. Taking common
denominator for all three rational numbers in equation one obtains just
the condition defining Pythagorean triangle!  This means that the
solutions of the equation can be written in the following form

\begin{eqnarray}
X-V&=& \epsilon_3  K \Delta (r,s, \epsilon_4) \nonumber\\
\frac{k}{l} &=& K\Delta (r,s,0)\nonumber\\
\frac{m}{n}&=& K\Delta (r,s, -\epsilon_4)\nonumber\\
\Delta (r,s,0) &= & (r^2+s^2)\nonumber\\
\Delta (r,s,1)&=& r^2-s^2\nonumber\\
\Delta (r,s,-1)&=& 2rs\nonumber\\
Y^2&=& 2V(-\epsilon_3  K \Delta (r,s,\epsilon_4)  +\epsilon_2 \Delta
(r,s,0)) \nonumber\\
c_2^2 &=& -\frac{b}{2}+ \epsilon_1 K\Delta (r,s,-\epsilon_4)
\end{eqnarray}

\noindent where $r,s$ are the integers defining Pythagorean triangle:
recall that one of these integers is even and one odd.  The result means
that solutions to the mass conditions are labeled by the rational number
$K$ and two integers.  The expression for $Y^2$ must reduce to one of the
following forms ($V<0$)

\begin{eqnarray}
Y^2&=& -2V K f\nonumber\\
f&=& 2r^2,2s^2 \ or \nonumber\\
f&=&(r+s)^2, (r-s)^2
\end{eqnarray}

\noindent This means that $Y^2$ is rational square  for $f=2r^2,2s^2$
 if $KV$  is rational square and  for $f=(r\pm s)^2$  if  $2KV$
is rational square.

\vm

It should be noticed also that the consistency condition allows rational
version
 of Lorentz group as symmetry group.  The quantity $(k/l)^2-(m/n)^2$
corresponds to the Minkowskian line element and Lorentz transformations
are realized as  $x\rightarrow \gamma (x-\beta y),  y\rightarrow \gamma
(y-\beta x)$, where both $\beta$ and  $\gamma= 1/\sqrt{1-\beta^2}$ are
rational numbers. This in turn implies that $\beta$ is in
one-one-correspondence
 with Pythagorean trangles! Pythagorean triangles define rational version
of 2-dimensional Lorentz group!  There is dual Lorentz group associated
with the side of length $2rs$. This group leaves invariant the
 purely nondiagonal form of line element (light cone coordinates and
transformations are represented
 as: $(r,s)\rightarrow (\frac{ \sqrt{1+\beta}}{\sqrt{1-\beta}}r,
\frac{\sqrt{1-\beta}}{\sqrt{1+\beta}}s)$.  In both cases the
transformations
 in general do not leave r and s integers and must be acocompanied by a
proper scalings.

 \vm

One can express the  angle parameter $s_1^2$ in terms of parameters
 $K,r,s$  by using the expressions of $X$ and $V$ in terms of $ s_1^2$

\begin{eqnarray}
s_1^2&=&\frac{ n_1}{n_1+n_2 + \epsilon_3 K\Delta (r,s,\epsilon_4)}
\end{eqnarray}

\noindent  From the requirement

\begin{eqnarray}
 s_1^2&\simeq& \frac{n_1}{9}\nonumber\\
n_1(d)&=&3(4)\nonumber\\
n_1(u)&=&5(6)
\end{eqnarray}

\noindent  guaranteing correct value for Cabibbo angle
  ($s_{Cab}\simeq  s_1(u)c_1(d)-c_1(u)s_1(d)$ )  one obtains (by noticing
that $n_1+n_2=11$ for both $U$ and $D$ case)

\begin{eqnarray}
K \Delta (r,s,\epsilon_4) &=& k-\delta  \nonumber\\
\epsilon_3&=&-1\nonumber\\
k&=& n_1+n_2-9
\end{eqnarray}

\noindent The parameters $\delta (u)$ and $\delta (d)$ are small
parameters,  which parametrize the deviation of mixing matrix from the
matrix describing mixing of two lowest generations only.

\vm

The expressions of $s_1$ and $s_3$ read as

\begin{eqnarray}
 s_1^2&=& \frac{n_1}{9+\delta }\nonumber\\
s_3^2&=& \frac{n_1+n_2-9-k+ \delta}{51} =\frac{\delta}{51}\nonumber\\
\
\end{eqnarray}

\noindent It should be noticed that $k<3$ is forced by the requirement
 that $s_3^2$ is positive.

 \vm

The expression for the cosine of CP breaking angle
is given by

\begin{eqnarray}
c_{CP}^2&=& \frac{Y^2}{R_0}\nonumber\\
Y^2&=& -2V( -k+\delta  +K\epsilon_1\Delta (r,s,0))\nonumber\\
R_0&=&\frac{ 4(u-n_1)(60-u)(u-9)}{u}\nonumber\\
V&=& 2n_2+n_1-69 +  2\epsilon_3K\Delta (r,s,\epsilon_4)\nonumber\\
u&=& n_1+n_2 -k +\delta = 9+\delta
\end{eqnarray}

\noindent From the expression of $u$ one find that $R_0$ is proportional
to  the small
 parameter $\delta$  and therefore $c_2^2$ becomes very large unless $R$ is
 also proportional to $\delta$ at this limit.  This amounts to the
following requirements

\begin{eqnarray}
K\Delta (r,s,0)&=& k-\delta_1\nonumber\\
\epsilon_1&=&1
\end{eqnarray}

\noindent which brings second small parameter in the theory and one can
write

\begin{eqnarray}
c_{CP}^2&=& \frac{2u}{4(u-n_1)(60-u)}(69-2n_2-n_1+2k-2\delta )
 (\frac{\delta_1}{\delta }-1) \nonumber\\
u&=& n_1+n_2 -k +\delta = 9+\delta
\end{eqnarray}

\noindent Unitarity requirement poses strong conditions on the ratio of the
 parameters $\delta$ and $\delta_1$. When the parameters are identical CP
breaking angles are vanishing and CKM matrix becomes purely real.

\vm

CP breaking angle is always nonvanishing as can be seen from the
expression  for the sine of the second angle of the  Pythagorean triangle

\begin{eqnarray}
sin(\phi_P)&=& \frac{\Delta (r,s,\epsilon_4)}{\Delta (r,s,0)}
= \frac{(k+\delta)}{(k+\delta_1)}
\end{eqnarray}

\noindent For small values of $\delta$ and $\delta_1$   one of the angles
of the triangle is very small.  Since the the two sides of  Pythagorean
triangle   are always even and odd numbers respectively it is not possible
to construct  Pythagorean triangle degenerate to line  so that
$\delta\neq  \delta_1$ holds always true. This in turn means that CP
breaking is a geometric necessity.

\vm

The value of $c_2^2$ can be expressed as

\begin{eqnarray}
c_2^2&=&  -\frac{b}{2}+\frac{\delta_2}{2} \nonumber\\
b&=& -\frac{Y^2+2VX}{Y^2+X^2}\nonumber\\
Y^2&=& -2V( \delta -\delta_1 )\nonumber\\
V&=& 2n_2+n_1-69-k+ \delta \nonumber\\
X&=& V-k+ \delta=  2n_2+n_1-69-2k+ 2\delta \\
K\Delta (r,s,0)&\equiv& \delta_2=
K\sqrt{\Delta (r,s,\epsilon_4)^2-\Delta (r,s,-\epsilon_4)^2}=\sqrt{
\delta^2-\delta_1^2+2k(\delta_1-\delta) }\nonumber\\
 \end{eqnarray}

\noindent The parameter $\delta_2$ is clearly not an independent quantity.

\vm

   The unitary constraints for $\delta$, $\delta_i$ can be written in
general
 form as

\begin{eqnarray}
1 &\leq &\frac{\delta_1}{\delta } \leq 1+  \frac{4(u-n_1)(60-u)}
{2u(69-2n_2-n_1- 2k+2\delta ) } \nonumber\\
0 &\leq& \delta_2 \leq 2+b\nonumber\\
u&=& n_1+n_2 -k +\delta = 9+\delta\nonumber\\
\end{eqnarray}

\noindent where we have not bothered to rewrite the definition of the
function b appearing in the  general expression of $c_2^2$.   For physical
solutions $c_2^2$ should be near unity, which corresponds to the upper
unitarity limit $\delta_2 = 2+b$.   In fact unitarity forces it to be
so.   The numerical  value of $-b/2$ is  slightly below  unity at the
limit $\delta,\delta_1  \rightarrow 0$ since one
 has

\begin{eqnarray}
-b (\delta,\delta_1 \rightarrow 0 )&\rightarrow& \frac{2(X+k)}{X}\nonumber\\
2\frac{( 69- 2n_2-n_1+k) }{( 69-2n_2-n_1+2k)    }&\le& 2
\end{eqnarray}

\noindent so that $s_2^2$ lies in rather narrow range
$(0,1-\frac{X+k}{X})$.

\subsubsection{Constraints on U and D matrices from empirical information
on  CKM matrix }

The most recent experimental information \cite{KMpaper} concerning CKM
matrix  elements
 is summarized in table below

\vl

\begin{tabular}{||l||} \hline \hline
$\vert V(1,3)\vert \equiv\vert V_{ub}  \vert =(0.087\pm 0.075)V_{cb}$ : $
 0.42 \cdot
10^{-3}<\vert V_{ub}\vert  <6.98\cdot 10^{-3}$   \\ \hline
 $ \vert V(2,3)\vert\equiv \vert  V_{cb}\vert = (41.2\pm 4.5)\cdot 10^{-3}$
 \\ \hline
$ \vert V(3,1) \vert \equiv \vert V_{td}\vert= (9.6 \pm 0.9) \cdot 10^{-3}$
 \\ \hline
$\vert V(3,2)\vert \equiv\vert V_{ts}\vert = (40.2 \pm 4.4)\cdot 10^{-3} $
\\ \hline
$s_{Cab}=0.226 \pm 0.002$\\ \hline
\end{tabular}

\vl

Table 4.4 \label{KMinfo} The experimental constraints on the absolute
values of the KM matrix elements.

\vm

 The expression for   Cabibbo angle given by

\begin{eqnarray}
sin(\theta_C)&\equiv& \sqrt{1-\vert V_{11}\vert^2}\nonumber\\
V_{11}&=&  c_1(u)c_1(d)+s_1(u)s_1(d)( c_3(u)c_3(d)+ s_3(u)s_3(d))\nonumber\\
&=& c_1(u)c_1(d)+s_1(u)s_1(d) cos(\theta_3(u)-\theta_3(d))\nonumber\\
&=& cos(\theta_1(u)-\theta_1(d))-s_1(u)s_1(d) (1- cos(\theta_3(u)-
\theta_3(d)))
\end{eqnarray}

\noindent  gives  for $\theta_3(u)-\theta_3 (d)=0$

\begin{eqnarray}
sin(\theta_c)&= &
\frac{(\sqrt{n_1(u)}\sqrt{9-n_1(d)}-\sqrt{n_1(d)}\sqrt{9-n_1(u)})}
{\sqrt{(9+\delta (d))(9+\delta
(u))}}
 \end{eqnarray}

\noindent  The requirement that the value is sufficiently near to the
observed  value $sin(\theta_c) \simeq 0.226 \pm 0.002$ gives constraints
on the values  $n_i(u/d)$ and $\delta (u/d)$.  \\ a) $(n_1(d)=3,n_2(u)=5)$
and  $n_1(d)=4, n_1(u)=6)$  yield same value at the $0.2236$  at  small
$\delta$ limit, which
 is just at the lower bound of experimental uncertainties. \\   b)
Nonzero value for  $\theta_3(u)-\theta_3(d)$ however tends to increase the
value of Cabibbo angle. If  $s_3(d)$ and $s_3(u)$ are of same sign maximum
of $s_{Cab}$  is achieved for $\delta (u)=0$  ( U matrix is nontrivial
for  two lowest generations only). \\ c)  For   $(n_1(d)= 4,n_1(u)=6)$
scenario the maximum value $s_{Cab} \simeq 0.22601$  (accidentally same as
experimental mean value!)  of  Cabibbo angle and is achieved at $\delta
(d)\simeq 0.032601$.\\ d)  For  $(n_1(d)= 3,n_1(u)= 5)$ scenario maximum
$s_{Cab}\simeq 0.2255$  is achieved at $\delta (d) \simeq  0.0031976 $. It
turns out that $\delta (u)=0$  implies too large values for the elements
$V(1,3)$ and $V(3,1)$. This decreases the prediction for Cabibbo angle so
that  the scenario $(n_1(d)=4,n_1(u)=6)$ is slightly favoured.  \\ e)  The
alternative $(n_1(d)=2,n_1(u)=4)$ yields at $\delta=0$ limit value larger
than experimental value and by choosing appropriately the values of
$\delta$  it is easy to reproduce the experimental value of  Cabibbo
angle.   The values of $\delta$ must be of order $1/2$ and are
suspiciously large.

\vm

The second constraint comes from smallness of $V(1,3)$ and $V(3,1)$.  From
the  table above it is clear that the moduli of the elements $V_{31}=
V_{td}$ and $V_{13}= V_{ub}$ should be
 below $ 10^{-2}$.  These elements correspond to the inner products of
first and third rows  for $U$ and $D$ type matrices.  By writing these
inner products explicitely one finds  for the imaginary part the expression

\begin{eqnarray}
Im(V(1,3))&=& c_2(u)s_1(d)sin(\theta_3(u)-\theta_3(d) ) s_{CP}(u)\nonumber\\
Im(V(1,3))&=& -c_2(d)s_1(u)sin(\theta_3(u)-\theta_3(d))  s_{CP}(d)\nonumber\\
\end{eqnarray}

\noindent  The values of these quantities are small if
 $s_{CP}(d),s_{CP} (u)$  and/or  $\theta_3(u)-\theta_3(d)$  are small
enough.
  $\theta_3(u)-\theta_3(d)$ is restricted from below by the requirement
that  Cabibbo angle  is within experimental uncertainties.

\vm

The real parts for $V_{13}$ and $V_{31}$  can be written
as

\begin{eqnarray}
Re(V(1,3))&=&
-c_1(d)(s_1s_2)(u)+ (s_1c_3)(d)( c_1s_2c_3+c_2s_3c_{CP})(u)\nonumber\\
&+&(s_1s_3)(d)( c_1s_2s_3-c_2c_3c_{CP})(u)
\nonumber\\
Re(V(3,1))&=&
-c_1(u)(s_1s_2)(d)+ (s_1c_3)(u)( c_1s_2c_3+c_2s_3c_{CP})(d)\nonumber\\
&+&(s_1s_3)(u)( c_1s_2s_3-c_2c_3c_{CP})(d)
\nonumber\\
\
\end{eqnarray}

\noindent  In order to minimize the values of $V(1,3)$ and $V(3,1)$   one
can vary the values of $\delta_1 (u),\delta_1 (d)$ and  choose the sign
for $s_2,c_2$ freely.
  Numerical experimentation shows that the correct manner to achieve
mimimization is to assume

\begin{eqnarray}
s_2(u), s_2(d) \ small \nonumber\\
c_2(d)>0,  c_2(u)<0
\end{eqnarray}

\noindent
The experimental bounds for the moduli of CKM matrix element give good
estimates for the parameters $s_1,s_2,s_3Ê$ appearing in the CKM matrix.

\begin{eqnarray}
s_1&=&.226 \pm .002\nonumber\\
s_1s_2&=& V(3,1) = (9.6\pm .9)\cdot10^{-3}\nonumber\\
s_1s_3&=& V(1,3)= (.0087 \pm .075)\cdot V(2,3)\nonumber\\
V(2,3)&=& (40.2 \pm 4.4) \cdot 10^{-3}
\end{eqnarray}

\noindent  The remaining parameter is $sin(\delta)$ or equivalently the
CP breaking parameter $J$

\begin{eqnarray}
J&=&Im( V(1,1) V(2,2) \bar{V}(1,2)\bar{V}(2,1))=
c_1c_2c_3s_2s_3s_1^2sin(\delta)\leq 6.7\cdot 10^{-5}\nonumber\\
\
\end{eqnarray}

\noindent where the upper bound is for $sin(\delta)=1$ and the previous average
values of the parameters $s_i,c_i$ (note that the poor knowledge of $s_3$
affects
on the upper bound for $J$ considerably).
  Information about the value of $sin(\delta)$ as well as on the range of
possible top quark masses  comes from CP breaking in $K-\bar{K}$  and
$B-\bar{B}$
systems.

\vm

 The observables in
 $K_L\rightarrow
2\pi$ system \cite{CPreview}

\begin{eqnarray}
\eta_{+-}&=& \frac{ A(K_L\rightarrow \pi^+\pi^-) }
{A(K_S\rightarrow \pi^+\pi^-)}=
 \epsilon+   \frac{\epsilon'}{1+\omega/\sqrt{2} }\nonumber\\
 \eta_{00}&=&
\frac{A(K_L\rightarrow \pi^0\pi^0)}  {A(K_S\rightarrow \pi^0 \pi^0)}=
\epsilon-2\frac{\epsilon'}{1-\sqrt{2}\omega}    \nonumber\\
\omega  &\sim& \frac{1}{20}       \nonumber\\
\epsilon&=& (2.27\pm .02)\cdot 10^{-3}\cdot exp(i43.7^o)\nonumber\\
\vert \frac{\epsilon'}{\epsilon}\vert &=& (3.3\pm 1.1)\cdot 10^{-3}
\end{eqnarray}

\nonumber  The phases of $\epsilon$ and $\epsilon'$ are in good approximation
identical.   CP breaking in $K-\bar{K}$   mass matrix
 comes  from the  CP breaking imaginary part of $\bar{s}d\rightarrow
s\bar{d}$
 amplitude  $M_{12}$ (via the decay to intermediate  $W^+W^-$ pair)
whereas  $K^0\bar{K}^0$ mass  difference $\Delta m$  comes from  the real
part of this amplitude: the calculation of the real part cannot be done
reliabely  for kaon since  perturbative QCD  does not work in the  energy
 region in
question.   On can however relate the  real part to the known mass difference
between
$K_L$ and $K_S$:   $2Re(M_{12})=\Delta m$.

\vm

     Using
the  results of  \cite{CPreview})  one can
express $\epsilon$  and $\epsilon'/\epsilon$  in the  following numerical
 form

\begin{eqnarray}
\vert \epsilon\vert& =&\frac{1}{\sqrt{2}} \frac{Im(M_{12}^{sd}}{\Delta
m_K}-.05\cdot  \vert \frac{\epsilon'}{\epsilon}\vert
=  2J (22.2 B_K \cdot
X(m_t) -5.6B'_K)\nonumber\\
\vert\frac{\epsilon'}{\epsilon}\vert&=&
2J\cdot 5.6 B'_K\nonumber\\
X(m_t)&=&\frac{ H(m_t)}{H(m_t=60 \ GeV)}\nonumber\\
H(m_t)&=&   -\eta_1 F(x_c)+\eta_2F(x_t)K
+\eta_3 G(x_c,x_t) \nonumber\\
x_q&=&\frac{m(q)^2}{m_W^2}\nonumber\\
K&=& s_2^2 +s_2s_3cos(\delta)\nonumber\\
\eta_1&\simeq& 0.7 \ \ \eta_2 \simeq 0.6 \ \ \eta_3 \simeq 0.5
\end{eqnarray}

\noindent  Here the  values of   QCD parameters $\eta_i$ depend on top mass
slightly  and are given for  $m_t=60  \ GeV$: $\eta_i$
as well as parameters appearing in the explicit expression of
$\epsilon'/\epsilon$
are assumed to be same for higher top masses in the following estimates.
  $B_K'$ and
$B_K$  are strong interaction matrix elements and vary between $1/3$ and $1$.
The
functions $F$ and $G$  \cite{CPreview}.
 are given by

\begin{eqnarray}
F(x)&=& x(\frac{1}{4}+\frac{9}{4}\frac{1}{1-x} -\frac{3}{2}\frac{1}{(1-x)^2})
+\frac{3}{2}(\frac{x}{x-1})^3 ln(x)\nonumber\\
G(x,y) &=&xy(\frac{1}{x-y}( \frac{1}{4}+\frac{3}{2}\frac{1}{1-x}
-\frac{3}{4}\frac{1}{(1-x)^2})ln(x) +(y\rightarrow x)
 -\frac{3}{4}\frac{1}{(1-x)(1-y)})\nonumber\\
\
\end{eqnarray}

\noindent  One can  solve parameter $B_K'$
by requiring that the value of $\epsilon'/\epsilon$ corresponds to the
experimental
mean value:

\begin{eqnarray}
B_K'&=& \frac{.05}{11.2\cdot J}\frac{\epsilon'}{\epsilon}
\end{eqnarray}

\noindent  One
can also solve the parameter $J$  and $sin(\delta)$ in terms of $\epsilon$
for a
given top mass.  Despite the poor knowledge of $V(1,3)$ one can make the
following
conclusions: \\ a) The requiment that $B_K'$ belongs to the range $(1/3,1)$
limits
the  values of CP invariant $J$ considerably:

\begin{eqnarray}
1.5\cdot 10^{-5}&\leq& J\leq 4.4\cdot 10^{-5}
\end{eqnarray}

\noindent
b) The requirement that $sin(\delta)$ is
nonnegative drops $k=89$ top from consideration: the point is that for large
 top
mass the sign of the function $H(m_t)$ changes and this implies that the
 relative
sign of $\epsilon$ and $\epsilon'$ changes. Also the absolute  value of $B_K'$
becomes larger than one.  \\ c) For $ k=97$ top quark one has
$sin(\delta)>1$. \\ d) For the  observed top quark  candidate one can
 reproduce all
values of $sin(\delta)$ in the allowed range for $J$  so that the precise
 value of
the  $CP$ breaking parameter remains open.

\vm

A second source of information comes from $B-\bar{B}$ mass difference. At the
energies in question perturbative QCD is expected to be applicable for the
calculation of the  mass difference and   mass difference is
predicted correctly if the mass of the top quark is essentially the mass of the
observed top candidate \cite{CKMdata}.  It seems  that top quark
cannot correspond to neither  to $k=89$ nor $k=97$  condensation level. A
possible
explanation for the discrepancy is that top quark corresponds to quantum
mechanical mixture of these levels and that the mass squared is therefore
expectation value: if top quark spends about $4$ percent of its
time on $k=89$ level the mass is the observed mass.

\vm

Physically acceptable  CKM matrices can be found numerically  by varying
 $\delta (u)$ and $\delta_1(d), \delta_1(u)$ inside unitary limits.  CKM
matrix is not
 sensitive  to the values of $n_2(d)$ and $n_2(u)$ for the physical
parameter values.
 The table below provides an example of $(n_1(d) =4,n_1(u)=6)$ CKM matrix
for which the values
 of matrix elements are within the experimental bounds. The value of
Cabibbo angle is  $s_{Cab}=0.22469$. The values of $s_2^2(d)$ and
$s_2(u)$  are rather small.    The value of CP-breaking parameter
$J=Im(V_{11}V_{22}\bar{V}_{12}\bar{V}_{21})$ is $J\simeq 4.986\cdot 10^{-5}$
quite near to the upper bound deduced above.
 It must be stressed that the example is  not probably the best one.  For
example, by suitably varying the parameters one could
 make $V_{td}$ smaller.

\vl

\begin{tabular}{||l|l|l|l||}\hline \hline
V&d&s&b\\ \hline
 u&9.7443 e-1 &       2.2459e-1-1.7011e-4i     & -1.5646e-3+ 6.4272e-3i    \\
\hline
 c&2.2445e-1 -8.0600e-5i &    -9.7368e-1-7.2035e-5i & 3.9319e-02 -
4.8791e-3i    \\
\hline
 t&5.1355e-3 +9.0110e-3i &        -3.6465e-2-1.3277e-2i        &
-9.6465e-1-2.6047e-1i   \\  \hline\hline
$\vert V\vert$ &d&s&b\\ \hline
u&9.7443e-1 &2.2459e-1 &.6.6149e-3\\ \hline
c&2.2445e-1& 9.7368e-1 & 3.9621e-2 \\ \hline
t&1.0372e-2 &3.8807e-2 &9.9919e-1\\ \hline\hline
\end{tabular}

\vl

Table 4.5.\label{KMexample} Example of   CKM matrix satisfying experimental
constraints \cite{KMpaper} with parameters  $n_1(d)=4,n2(d)=11$,
$n_1(u)=6,n_2(u)=6$,
 $\delta (d)=0.0318 $,
 $s_2^2(d)=8\cdot 10^{-5}$,
 $\delta (u)=0.009 $,  $s_2^2(u)=7\cdot 10^{-4}$. The value of CP breaking
invariant $J$ is $J=4.986\cdot 10^{-5}$.

\subsubsection{Number theoretic conditions on U and D matrices}

KM matrix  represents together with quark and lepton masses  the   basic
parameters of the  standard model. It would be nice if p-adic approach
would predict only the masses  but also  U  and D matrices uniquely.
The p-adic and real probability concepts  are formally equivalent if the
moduli  squared of U,D and
 CKM matrices are rational numbers. This requirement indeed poses strong
number
 theoretical conditions on U and D matrices and gives hopes of getting
unique CKM matrix.

\vm

Rational moduli squared  are obtained if the elements of the U and D
matrices  are rational numbers.
 This implies that the angles appearing in the matrices correspond to
Pythagorean
 triangles.
 It is not however clear  whether the Pythagorean conditions can be
satisfied.
 This forces to look carefully for the general structure of U,D and CKM
matrices. The following
 observations emerge. \\ a)  In p-adic  regime, what might  be called
irreducible phases,  are possible.
 For instance, $x=\sqrt{5}$ does not exist as rational number but one can
replace it with  rational quantity  $2+i$  having irreducible phase, which
cannot be cancelled away in the realm of rationals.  Also irrational
phases are possible: $\sqrt{3}\rightarrow \sqrt{2}+i$ gives simple example
of this phenomenon.   If one can replace ordinary square roots with square
roots containing irreducible phase, which is  constant along rows of U(D)
matrix one can perhaps obtain unitary and rational U (D) matrix. Even
irrational irreducible phases are possible if only moduli squared  are
required to be rational.\\ b) The study of U (D)  matrix shows that
$c_1,s_1$ and $c_{CP},s_{CP}$ cannot have irreducible phase: the reason is
that these quantities do not appear homogenously on the rows of $U/D$
matrix. It also turns out that the Pythagorean conditions do not lead to
any contradictions.  \\
 c) The parameters $s_2,c_2$ can  be given common irreducible phase
without affecting the rationality of the moduli squared  or  spoiling
unitarity conditions. The phase
 cancels also from the moduli squared  of CKM matrix elements.   A nice
manner to get common  irreducible rational phase  is to assume that these
angles correspond to
 triangle with integer valued shorter sides and  to require that the
square root  $\sqrt{X^2+Y^2}$ of  their  common
 denominator $\sqrt{X^2+Y^2}$   is  replaced with $X+iY$.    Number
theoretic constraints come from the requirement that $\frac{s_2^2}{c_2^2}$
is square of a rational number. \\ d) Also $s_3,c_3$ can be given same
phase same for all
 rows.   U and D matrix unitarity remain unaffected and probabilities
rational.  This phase  must cancel from KM matrix and this requires that
the phase is identical for D and U matrices.
  The general form of $s_ 3^2= \frac{\delta}{51}=\frac{km}{kn51}$
suggests  that the possible  phase comes from the denominator $kn51$:
$\sqrt{kn51}\rightarrow a+ib$. The phase  can  be chosen to be rational
and $s_3,c_3$  correspond to a triangle with integer valued shorter sides.
  Pythagorean triangle  is possible for the scenario solving proton spin
crisis, only.
  The phases associated with U and D matrices are identical if
$k(U)n(U)/k(D)n(D)$ is square of a rational number.

\vm

To sum up, the most general scenario allowing rational probabilities
satisfies
 following  conditions

\begin{eqnarray}
&s_1,c_1& \  Pythagorean\nonumber\\
&s_{CP},c_{CP}& \ Pythagorean\nonumber\\
&\frac{s_2}{c_2}& \ rational \nonumber\\
&\frac{s_3}{c_3} &\ rational \nonumber\\
&\frac{n(U)k(U)}{n(D)k(D)}=(\frac{k}{l})^2
\end{eqnarray}

\noindent   For rational U,D and CKM matrices all these angles are
Pythagorean. In the following the number theoretical consistency of
 these conditions is studied.  The method is to write various conditions
 modulo 8 (modulo n for arbitrary n gives valuable information).  The
reason is that the square of an
 odd number is always one modulo 8. In particular,  the squares for the
sides  of Pythagorean  triangle are equal to $1$ and $0$ modulo 8 so that
modulo 8  equations are easily derivable.  The  ratio $s^2/c^2$ for
Pythagorean angles  ($s_1, c_1)$ and $s_{CP},c_{CP}$)   imply congruence
relating the parameter $K$ to the parameters $r,s$ and integers $n_i,k$.
Similar congruences are obtained from the requirement that $s_2/c_2$ is
rational for $s_2,c_2$ and $s_3,c_3$.  Four independent nonlinear
congruences are obtained as consistency conditions for U and D matrix so
that  these matrices might be even unique!

\vm

\begin{center}
{\bf  1. Pythagorean conditions force the scenario solving
proton spin crisis}
\end{center}

\vm

  The most stringent requirement  on U and D matrix is that  even the
sines and cosines are rational numbers and therefore correspond to
Pythagorean squares $(r,s)$ with sides $r^2-s^2$ and $2rs$ ( $r$ or $s$
are not both even or odd) and sines and cosines  are given by  the
expressions

\begin{eqnarray}
sin(u)&=& \frac{2rs}{(r^2+s^2)}\nonumber\\
cos(u)&=& \frac{( r^2-s^2)}{(r^2+s^2)}
\end{eqnarray}

\noindent  or by the expression obtained by the
 exchange $2rs\leftrightarrow r^2-s^2$.

\vm

To see whether  the mass conditions  allow all angle parameteres to
be Pythagorean angles  one can write the equations modulo 8 and use the
elementary fact the square of odd number is always one modulo
8 whereas the square of even number is always $4$ or $0$ modulo 8. As a
consequence one has

\begin{eqnarray}
 (r^2+s^2)^2=(r^2-s^2)^2&=& 1 \  mod  \  8 \nonumber\\
(2rs)^2= 0 \ mod \ 8
\end{eqnarray}

\noindent The equations  for $D$,  when written modulo $8$ give

\begin{eqnarray}
 \vert D_{12}\vert ^2 + 4 \vert U_{13}\vert ^2 &=& n_1(d)\nonumber\\
\vert C_{22}\vert ^2 + 4\vert D_{23}\vert ^2 &=& n_2(d) \nonumber\\
\end{eqnarray}

\noindent Same applies to $U$ matrix.  By writing the left hand side of the
equations explicitely in terms of the parametrization one finds that  the
first term on the  left hand side of the equations  is sum of one or two
terms with value equal to 0 or 1 and therefore can have only the values
$0,1,2$.   This implies that left hand side can  have value in the set $
n= 0,1,2,4,5,6, 7$, which does not contain $n=3$.   Therefore   $n_1(d)=3$
is impossible to realize using rational $D$ matrix whereas   $n_1(d)=4$
favoured by proton spin crisis could allow rational U and D matrices.

\vl

\begin{center}
{\bf 2.  Pythagorean conditions for $(s_1,c_1)$}
\end{center}

\vm

The Pythagreaon conditions for $(s_1,c_1)$ read as

\begin{eqnarray}
s_1^2&=& \frac{\Delta (r_1,s_1,\epsilon (1))}{k_1}\nonumber\\
c_1^2&=& \frac{ \Delta (r_1,s_1, -\epsilon (1)}{k_1} \\
\end{eqnarray}

\noindent where the functions $\Delta$ are already familiar. Using the
expression $s_2^2=n_1/(9+\delta)$ one finds that  the Pythagorean triangle
condition  for $s_1^2$ gives the following
 possibilities

\vl

\begin{tabular}{||c|c|c||}\hline\hline
  $n_1k_1$  & $((9-n_1)n+m)k_1$   & $(9n+m)k_1$ \\ \cline{1-3}\hline
 $(2r_1s_1)^2$  & $( r_1^2-s_1^2)^2$&$(r_1^2+s_1^2)^2$  \\  \hline
  $(r_1^2-s_1^2)^2$  & $(2r_1s_1)^2$&$(r_1^2+s_1^2)^2$\\  \hline
\end{tabular}

\vl

Table 4.6.  \label{Possib} Various alternative solutions to Pythagorean
 conditions for
$(s_1,c_1)$

\vm

\noindent Here $k_1$ is an rational number  resulting from the
nonuniqueness of the representations $\delta = m/n $. There
 are two cases corresponding to odd and even $n_1$. The condition
$n_1nk_1=1 \ mod \ 8$ holds true as difference
 of sides squared
 for Pythagorean triangle. The condition implies is satisfied if
$n_1k_1=1 \ mod \ 8$ and $n=1 \ mod \ 8$  holds true, which leaves only
the second alternative for the sides of Pythagorean triangle.
 This in turn implies $(9n+m)k_1=1 \ mod \ 8$, which requires  $9n+m$ to
be  divisible by  $n_1$ and
 $(9n+m)/n_1=1  \ mod \ 8$.  This gives

\begin{eqnarray}
n&=& 1+ 8p\nonumber\\
m&=& n_1 -1 + (qn_1-9p-1)8\nonumber\\
\delta &=& \frac{ n_1 -1 + (qn_1-9p-1)8       }{1+8p}
\end{eqnarray}

\noindent The two conditions for $\delta$ do not lead to any obvious
contradictions.

\vl

Pythagorean condition allow to solve $K$  in terms of $(r,s,r_1,s_1)$ from
the ratio $s_1^2/c_1^2$, which does not contain  $k_1$.

\begin{eqnarray}
K&=& n_1-9+ n_1\frac{\Delta^2 (r_1,s_1,-\epsilon (1) )}{\Delta^2
(r_1,s_1,
\epsilon (1)}.
\end{eqnarray}

\noindent Later an analogous formula will be derived from $s_{CP},c_{CP}$
 Pythagorean conditions and this gives rise to consistency condition
between various integers.

\vl

\begin{center}
{\bf 3.   $s_3,c_3$ can be Pythagorean}
\end{center}

\vl

 The moduli squared of  U and D  matrix  elements are  rational if  the
shorter sides of the $s_3,c_3$ triangle can be chosen to be integers. Here
it   will be shown that the conditions for
 rational phase for  $s_3,c_3$ are in accordance  with the condition $n=1
\ mod \  8$ derived from $s_1,c_1$ conditions and that also Pythagorean
alternative is possible.  For $s_3^2= \delta /51$ and $c^2_3$ the  rational
 irreducible phase  conditions imply that these angles correspond to a
triangle with integer valued shorter sides $k_3,l_3$ and the possible
rational phase can be chosen to be  $k_3+il_3$

\begin{eqnarray}
\delta&=& \frac{ km}{kn}\nonumber\\
51k n&=& k_3^2+l_3^2\nonumber\\
51kn&=&m^2+r^2
\end{eqnarray}

\noindent     The condition  $51kn=k_3^2+l_3^2 \ mod \ 8$  together with
$n= 1 \ mod \ 8$ implies $3k= k_3^2+l_3^2 \ mod \ 8$, which gives the
  following possibilities  for $k$  and $k_3$ and $l_3$

\vl

\begin{tabular}{||l|l|l|l|l|l||}\hline\hline
 $k  \ mod  \ 8 $&0&3&6&4&7\\ \hline
$k_3^2+l_3^2 $& 0   &1&2&4&5   \\  \hline\hline
\end{tabular}

\vl

Table 4.7. \label{Kvalues} The values of $k \  mod \  8$ for various
combinations
 of $k_3,l_3$.

\vm

The ratio $s_3/c_3=(k_3/l_3)$ is rational and this gives linear
congruence  relating $K$ to the parameters $(r,s)$ and rational number

\begin{eqnarray}
K&=& \frac{(51 \beta_3^2 -k(1+\beta_3^2)}
{\Delta (r,s,\epsilon_4)(1+\beta_3^2)}
\nonumber\\
\beta_3&=&\frac{k_3}{l_3}
\end{eqnarray}

\noindent If $s_3,c_3$ corresponds to Pythagorean triangle the additional
condition $\beta= (r^2-s^2)/2rs$ or $\beta = 2rs/(r^2-s^2)$ holds true.

\vm

The requirement that the possible irreducible phases associated with $U$
and $D$ matrices are identical implies the condition

\begin{eqnarray}
\frac{k(U)n(U)}{k(D)n(D)}&=&\gamma^2
\end{eqnarray}

\noindent where $\gamma$ is rational number. For Pythagorean case this
condition is not needed.

\vl

\begin{center}
{\bf  4.  Pythagorean conditions for $c_{CP},s_{CP}$}
\end{center}

\vm

$c_3,s_3$ should be rational numbers in order to get rational moduli
squared for U,D and  CKM matrices.  The general
 expression for $c_{CP}^2$ reads as

\begin{eqnarray}
c_{CP}^2&=&\frac{Y^2}{R_0}\nonumber\\
Y^2&=& 2(72-2n_2-n_1-2\delta ) ( \delta_1 -\delta)\nonumber\\
R_0&=& 4(51c_1 abs(c_3s_3)  )^2=\frac{ 4(u-n_1)(60-u)(u-9)}{u}\nonumber\\
u&=&9+\delta \nonumber\\
\delta &=& \frac{ n_1 -1 + (qn_1-9p-1)8       }{1+8p}
\end{eqnarray}

\noindent  $R_0$ contains only $abs(c_3s_3)$ term so that p-adically
rational  square root exists for $R_0$.  For $Y^2$ square root exists also
provided the  function $-2Vf$
 is rational square ( $Y^2=-2VKf$ with  $f=2r2,  2s^2$ or  $f=(r\pm s)^2$).
Therefore only the rationality  condition for $s_{CP} $ can  turn out
troublesome.
 The common factor $2$ drops from the ratio defining $c_{CP}^2$.

\vm

  The Pythagorean conditions read in present case as

\begin{eqnarray}
Y^2 k_{CP} &=& (r_{CP}^2-s_{CP}^2)^2\nonumber\\
R_0k_{CP}&=& (2r_{CP}s_{CP} )^2
\end{eqnarray}

\noindent Also the second alternative obtained by permuting the sides of
 the triangle turns out to be excluded.  $k_{CP}$ is essentially the common
denominator of rational numbers involved.

\vm

Again one can write the conditions modulo 8 in order to derive
consistency conditions.
The general form for $Y^2$ and $R_0$  modulo 8 are

\begin{eqnarray}
Y^2/2&=& -((2(n_1+n_2)+3)(-\delta_1 -n_1+1)=-(5+2k)(-\delta_1 -n_1+1)
\nonumber\\
R_0/2&=& \frac{32(4-n_1)(n_1-1)}{n_1}\nonumber\\
\frac{Y^2}{R_0}&=& \frac{ -(5+2k)(-\delta_1 -n_1+1) n_1   }
{ 32(4-n_1)(n_1-1)         }\nonumber\\
k&=& n_1+n_2-9
\end{eqnarray}

\noindent   From this expression it is clear that for even $n_1$ one can
always divide $R_0$ by $2n_1$  and still have   $R_0/2n_1=0 \ mod \ 8$ for
interesting values of even $n_1$.  The division with  $n_1$ implies  the
choice $k_{CP}=k_B/2n_1$  in general so that one has

\begin{eqnarray}
Y^2k_{CP}&=& (5+2k)(-\delta_1+n_1-1)k_B=1 \ mod \ 8
\end{eqnarray}

\noindent Writing $\delta_1= p_1/q_1$ and taking $k_B=q_1$ one has

\begin{eqnarray}
-p_1+(n_1-1)q_1&=&s(k) \ mod \ 8  \nonumber\\
s(k\equiv n_1+n_2-9=2)&=& 1 \ mod \ 8 \nonumber\\
s(k=1)&=& 7 \ mod \ 8
\end{eqnarray}

\noindent This condition does not  lead to no obvious contradiction with
earlier
 results.

\vm

The explicit expression of $Y^2$  and $R_0$ allows to write the
corresponding  Pythagorean conditions in explicit form

\begin{eqnarray}
 2(69- 2n_2-n_1 +K\Delta (r,s,\epsilon_4))Kf&=& (r_{CP}^2-s_{CP}^2)
\nonumber\\
\frac{4(u-n_1)(60-u)(u-9)}{u}k_{CP}&=&(2r_{CP}s_{CP})^2\nonumber\\
f &=& 2r^2,2s^2 \ or \  (r\pm s)^2 \nonumber\\
u&=& n_1+n_2-K\Delta (r,s,\epsilon_4)
\end{eqnarray}

\noindent   Taking the ratio of the conditions one obtains equations given
$K$
 in terms of the integers associated with two  Pythagorean triangles and
integers $n_i,k$.

\begin{eqnarray}
2(&69&- 2n_2-n_1 +K\Delta (r,s,\epsilon_4))Kf(r_{CP}s_{CP})^2u
\nonumber\\
 &=&
 (u-n_1)(60-u)(u-9)(r_{CP}^2-s_{CP}^2)\nonumber\\
u&=& n_1+n_2-K\Delta (r,s,\epsilon_4)
\end{eqnarray}

\noindent The resulting equation is third order in $K$ and one must
require that
 the solutions are rational, which gives additional conditions for the
integers $r,s$ and  $r_{CP},s_{CP}$ appearing in the equation.  One can
also subtitute the expression for $K$ obtained  from the Pythagorean
conditions for $s_1$ to this equation and get nonlinear congruence as a
consistency
 condition relation for the integers $r,s,r_1,s_1,r_{CP},s_{CP}$.

 \vl

\begin{center}
{\bf 5. Irreducible phase conditions for $c_2,s_2$}
\end{center}

\vm

The remaining number theoretic conditions come from $c_2^2$ and $s_2^2$.
 There are two possible
 manners to obtain rational probabilities. \\ a) $c_2$ and $s_2$ are
rational numbers (Pythagorean triangle)  so that  phases are trivial.\\
b)   $c_2$ and $s_2$ allow identical  irreducible rational phases coming
from the common denominator of the quantities defining these parameters.
 \\
  The analysis of consistency  conditions proceeds by writing
$c_2^2/s_2^2$ as rational square and by looking for the consequences.
 Also modulo $8$ analysis is possible.

\vm

  The expressions for $c_2^2$ and $s_2^2$ read as

\begin{eqnarray}
s_2^2 &=& \frac{ Y^2(1-\delta_2)+ X(-2V +X(2-\delta_2))}{2(Y^2+X^2)}
\nonumber\\
 c_2^2&=&  \frac{ Y^2(1+\delta_2)+ X(2V +X\delta_2)}{2(Y^2+X^2)}
\nonumber\\
X&=& 2n_2+n_1-69-2k+2\delta\nonumber\\
V&=& X+k-\delta\nonumber\\
\end{eqnarray}

\noindent  The obvious manner to get irreducible phase is to perform  the
replacement $X^2+Y^2\rightarrow X+iY$. One must however be careful since
$Y^2$ and $X^2$ are only formally squares and it might be necessary to
introduce
 rational  multiplier for $X^2+Y^2$ to get irreducible phase.

\vm

 The requirement is that the numerators of $s_2^2$ and $c_2^2$ are rational
numbers implies that the ratio $c_2^2/s_2^2$ is square of a rational
number:

 \begin{eqnarray}
\frac{ Y^2(1+\delta_2)- X(2V +X\delta_2)}
{ Y^2(1-\delta_2)+ X(2V +X(2-\delta_2)) }&=&\alpha^2\nonumber\\
\alpha&=& \frac{k_2}{l_2}
\end{eqnarray}

\noindent  Here the rational number $\alpha$ is restricted to rather
narrow  range by unitary. This gives second order equation for $X $ with
consistency conditions resulting  from the requirement that the argument
of the square root appearing in the solution is square of  a  rational
number.

\begin{eqnarray}
X &=& -\frac{b}{2}- \sqrt{ b^2-4c}\nonumber\\
b&=&  2\frac{(   (\delta_1-\delta) ((1+\delta_2)-(1-\delta_2)\alpha^2)+
(k-\delta_2)(1+\alpha^2))     } {  (2+\delta_2)(1+\alpha^2) } \nonumber\\
c&=&    -\frac{2(\delta_1-\delta) (k-\delta)( 1-\alpha^2+\delta_2
(1+\alpha^2)) } { (2+\delta_2)
(1+\alpha^2)  } \nonumber\\
 \alpha&=& \frac{k_2}{l_2}
\end{eqnarray}

\noindent The consistency condition for  this equation is that $b^2-4c$ is
square  of a  rational number.

\begin{eqnarray}
b^2-4c&=& \beta^2\nonumber\\
\beta&=& \frac{k_3}{l_3}
\end{eqnarray}

\noindent For Pythagoran triangle one must replace $\beta$ with the ratio
of
 shorte sides of Pythagorean triangle. It is possible to perform modulo 8
analysis for this
 condition using the information about f $\delta$ and $\delta_1$ (recall
that $\delta_2$ is  expressible in terms of $\delta_1$ and $\delta$). In
particular, one can feed in  the modulo 8   information about $\delta$ and
$\delta_1$.

\vm

 Since the quantities $\delta,\delta_1,\delta_2$ are all defined in terms
of
 the parameters $K,r,s$ defining the basic Pythagorean triangle  the
consistency
 condition gives condition for  these three numbers plus integers $n_i$.
Using the definitions

\begin{eqnarray}
K\Delta (r,s,\epsilon_4)&=&k-\delta \nonumber\\
K\Delta (r,s,0) &=& k+\delta_1\nonumber\\
K\Delta (r,s,-\epsilon_4) &=& \delta_2
\end{eqnarray}

\noindent one can transform this condition to fourth order equation for
$K$.
  One manner to proceed is to solve this consistency condition for $K$.
 The rationality requirement  for the solution can be used  also now to
derive constraints on the integers $r,s$ of allowed
 Pythagorean triangles.    A second manner to proceed is to substitute the
expression of  $K$ in terms of $r_1,s_1,r,s$ to the equation and solve the
resulting  nonlinear congruence.
 The requirement that integer solution is obtained is quite restrictive.

 \subsubsection{Summary}

It is useful to collect the results found from the number theoretic
analysis.\\ a) In principle it  is   possible to obtain U,D and CKM
matrices defining rational  probabilities by allowing common rational
irreducible phases for $s_2,c_2$ and $s_3,c_3$ respectively whereas  $s_1$
and $c_1$  and $s_{CP},c_{CP}$ are rational numbers and correspond to
certain Pythagorean triangles. U and D matrices can be also rational for
the scenario solving proton spin crisis so that  all angles correspond to
Pythagorean triangles. Rational $su(3)$  matrices form a group and probably
 the mathematical literature
contains explicit characterization of the matrix elements:  it might be easy
to see how uniquely the two  constraints
for mass squared expectation values fix the physically
allowed $U$ and $D$ matrices.
  \\
   c) The requirement that the tangents associated with the four angle
parameters are rational implies four consistency conditions allowing to
express  the parameter $K$ as a solution of, in general nonlinear,
rational congruence.
 Thus there are altogether $4$ different equations for the parameter $K$
and these must be mutually consistent! If Pythagorean triangles are in
question the expressions for the tangents of the angles must satisfy
additional conditions.
That only few solutions to congruences exist is suggested by classical result
 of number theory
\cite{Chowia}:
the congruence $P(x,y)=c$, where $P$ is homogeneous polynomial  of $x,y$
 with integer  coefficients of degree $n\ge 3$  and $P(x,1)$ is
 irreducible in field of rationals has only finite number of
integer solutions.
 \\ d) There is also a consistency condition relating $U$ and  $D$
matrices, which
 comes from the requirement that the irreducible phases associated with
$c_3,s_3$ are identical
 for U and D.
  \\ d) Unitarity gives bounds on the parameters
$\delta,\delta_1,\delta_2$: in particular
 $s_2$ varies in rather narrow range $(0,1-(X+k)/X)$ as a consequence.
 The only possibly Pythagorean  alternative  is  $(n_1(d)=4
,n_2(d)=11,n_1(u)=6,n_2(u)=6)$ scenario,  which  provides
  good
  understanding of  hadron masses,  allows identical mixing matrices for
baryonic and mesonic quarks, allows nonvanishing first order color binding
 energy  also for baryons and
 provides a possible solution to the proton spin crisis. $K-\pi$ mass
difference  favours $n_2(d,M)=11<n_2(d,B)=13$ for mesons: CKM matrix is
essentially identical with  the baryonic one also with this choice.   CP
breaking is a geometric necessity resulting from the
 impossibility of   Pythagorean triangle with one  side  having  zero
length.   For a  suitable
 choice of free parameters the elements of  CKM matrix are within
experimental bounds and CP breaking parameter
 $\epsilon$ has correct order of magnitude.

\subsection{ Color Coulombic interaction, color magnetic
hyperfine splitting
and isospin-isospin interaction }

Color coulombic interaction gives contribution, which can is in principle
 different for each quark pair since the average quark distances need not
be identical so that one has

\begin{eqnarray}
\Delta s_c&=& \sum_{pairs} c(q(i),q(j))
\end{eqnarray}

\noindent where the elements $c(q(i),q(j))$ are  negative  integers.  In
 the sequel the elements  are assumed to be constant for each quark pair
but different for baryonic  and mesonic quarks.

\vm

Color magnetic hyperfine splitting  makes it possible to understand
 the $\pi-\rho$,$K-K*$, $N-\Delta$, etc.
 mass differecences \cite{Close}.  The interaction energy is of form

\begin{eqnarray}
\Delta E&=& S\sum_{pairs} \frac{\bar{s}_i\cdot\bar{s}_j}{m_im_jr_{ij}^3}
\end{eqnarray}

\noindent The effect is so large that it must be p-adically first order
and  the generalization of the mass splitting formula is rather obvious:

\begin{eqnarray}
\Delta s&=& \sum_{pairs} S(i,j)\bar{s}_i\cdot \bar{s}_j
\end{eqnarray}

\noindent The coefficients $S_{ij}$ depend must be such that integer
 valued $\Delta s$ results and Planck masses are avoided: this makes the
model highly predictive.  Coefficients   can depend both on quark pair and
on hadron since the size of hadron need not be constant. In any case,
very limited  range of possibilities remains for the coefficients.

\vm

In principle also interaction between electroweak isospins is possible in
hadron.
 The understanding of the mass of $\Omega$ seems  to be difficult unless
this
 interaction is  present.   This interaction seems to occur mainly between
electoweak isospins of  fermions of same generation so that the
corresponding symmetry group is tensor power of  electroweak $su(2)$
groups rather than some larger group as in GUTs.  This implies a
deviation from
 age old  $su(3)$ picture of light hadrons.

\vm

Some examples are useful in order to make new picture familiar.
  For instance for $uds$ baryon   $ud$ pair belongs to irreducible
electroweak
 multiplet with $J=1$ or $J=0$ and $s$ is doublet. For $ssd$ type baryon
$ss$ pair is in $J=1$ or $J=0$ multiplet. For $sss$ state (the troublesome
$\Omega$) isospins form state in $I=3/2$  multiplet.  The remainging
states ($ccc,ccs,css$) of this multiplet contain charmed quarks. $ssu$ and
$ssd$ ($\xi$)  baryons belong to $I=0$ strange multiplet.

\vm

The general form of the electroweak  isospin-isospin interaction is same
as of
  color   magnetic interaction. Each generation  pair $(g_1,g_2)$,
$g_i=0,1,2$ can have
 different  interaction strength $K(g(i),g(j))$.

\begin{eqnarray}
\Delta s&=&  \sum_{pairs } K(g(i),g(j))\bar{I}_i\cdot \bar{I}_j\nonumber\\
\
\end{eqnarray}

\noindent and no Planck mass requirement gives constraints on the values of
 the matrix $K$.

\subsubsection{Baryonic case}

Consider first the determination of $S(i,j)$ and $K(g(i),g(j))$ in case of
baryons.
 The general splitting pattern  for baryons resulting from color
Coulombic,
 spin-spin and isospin-isospin interactions is given by the following
table.  It should be noticed that  $\Sigma$ baryons are not eigenstates of
total electroweak isospin but superpositions of $I=3/2$ and  $I=1/2$
states with equal  weights.  Furthermore, $\Omega$ is $I=3/2$ state and
the isospin of $ss$  pair in $\Xi$ is $I_{12}=1$.

\vl

\begin{tabular}{||l|l|l|l|l|l|l|l||}\hline\hline
baryon& J&$J_{12}$&I&$I_{12}$& $\Delta s^I$ &$\Delta s^{spin}$
 &$s_{eff}$\\
 \hline
N&$\frac{1}{2}$&0&$\frac{1}{2}$&0& $-\frac{3}{4}K(0,0) $&
$-\frac{3}{4}S(d,d)$
& 18\\ \hline
$\Delta$&$\frac{3}{2}$&1&$\frac{3}{2}$&1& $\frac{3}{4}K(0,0)$&
 $\frac{3}{4}S(d,d)$      &31\\ \hline
$\Lambda=uds$&$\frac{1}{2}$&0&$\frac{1}{2}$&0& $
-\frac{3}{4}K(0,0)
 $&-$\frac{3}{4}S(d,d) $     &25\\ \hline
$\Sigma =uds $&$\frac{1}{2}$&0&$\frac{1}{2} \oplus \frac{3}{2}$&1
&$\frac{1}{4}K(0,0)-\frac{1}{4}K(0,1)$ &-$\frac{3}{4}S(d,d)$ &29\\ \hline
$\Sigma^* $&$\frac{1}{2}$&0&$\frac{1}{2} \oplus \frac{3}{2}$& 1&
$\frac{1}{4}K(0,0)-\frac{1}{4}K(0,1)$ &$\frac{1}{4}S(d,d)+
\frac{1}{2}S(d,s)$
&40\\ \hline
$\Xi =ssd$ &$\frac{1}{2}$&0&$\frac{1}{2}$&1&
$\frac{1}{4}K(1,1)-K(0,1)$&-$\frac{3}{4}S(s,s)$   &36\ \\ \hline $\Xi^*$
&$\frac{1}{2}$&0&$\frac{1}{2}$&1&$-\frac{3}{4}K(1,1)$&$\frac{1}{4}S(s,s)+
 \frac{1}{2}S(d,s)$   &48\
\\ \hline $\Omega $&  $\frac{3}{2}$&1&$\frac{3}{2}$&1&
$ \frac{3}{4}K(1,1) $
&$\frac{3}{4}S(s,s)$ &58\\
\hline \hline \end{tabular}

\vl

Table 4.8. \label{Isobsplit} Color Coulombic, spin-spin and isospin-isospin
splittings for baryons.

\vm

\noindent Spin-spin and isospin-isopin  splittings  are deduced from the
formulas

\begin{eqnarray}
\Delta s^{spin}&=& S(q(1),q(2)) (\frac{J_{12}(J_{12}+1)}{2}-\frac{3}{4})
\nonumber\\
&+&
\frac{1}{4}(S(q(1),q(3)) +S(q(2),q(3)))(J(J+1)-J_{12}(J_{12}+1)-
\frac{3}{4})
\nonumber\\
\Delta s^{I}&=& K(g(1),g(2)) (\frac{I_{12}(I_{12}+1)}{2}-\frac{3}{4})
\nonumber\\ &+&
\frac{1}{4}(K(g(1),g(3)) +K(g(2),g(3)))(I(I+1)-I_{12}(I_{12}+1)-
\frac{3}{4})
\nonumber\\
\
\end{eqnarray}

\noindent where $J_{12}$ is the angular momentum eigenvalue of the 'first
two
 quarks', whose  value is fixed by the requirement that magnetic moments
are of correct sign.
 All baryons are eigenstates of $I$ and $I_{12}$  except $\Sigma=uds$,
where
 $ud$ is $I_{12}=1$ state so that state is superposition of $I=1/2$ and
$I=3/2$ states with equal weights.

\vm

The masses determine the values of the parameters uniquely if one assumes
that color binding  energy is constant:  $c(q(i),q(j))=c$  Assuming that
the values of $s(eff,q(i))$ are $s(eff,d)=s(eff,u)= 9$ and $s(eff,s)=16$
one obtains  the following table

\vl

\begin{tabular}{||l|l|l|l|l|l|l||}\hline \hline
$K(0,0)$&$K(0,1)$&$K(1,1)$&$S(d,d)$&$S(d,s)$&$S(s,s)$&   $c$\\   \hline
$4$&$0$&$10$ &$5-\frac{1}{3}$&$12+\frac{2}{3}$&$7-\frac{1}{3}$&$-1+
\frac{1}{6}$ \\ \hline
$4$&$0$&$10$ &$4$&$12$&$6$&$-1$ \\ \hline\hline
\end{tabular}

\vl

Table 4.9. \label{Splittingbparam} The values of the parameters
characterizing  baryonic spin-spin and isospin splittings of baryons in
order $O(p)$. The lower row gives  the best integer values for the
parameters.

\vm

\noindent  These parameter values in the table  produce the first order
 contributions exactly. If one requires that the parameters are integers
lead to errors in mass prediction below two per cent.
 Remarkably, the values of $K(i,j)$ are integers.  $K(0,1)$ vanishes as in
bosonic case, where the vanishing is implied by the  smallness of
 CP breaking and  this suggests than in first order the  isospin-isospin
interaction between  different  generations vanishes: a possible
explanation is that exchange force is in question.

\vl

\begin{tabular}{||l|l|l|l|l|l|l|l|l||}\hline\hline
baryon& N& $\Delta$ &$\Lambda$& $\Sigma $ & $\Sigma^* $ &$\Xi$ &$\Xi^*$&
 $\Omega $ \\ \hline
$ s_{eff}$ &   18  &30   &25  & 29 &40&36&48&58 \\ \hline
$s_{eff}(exp)$&18  &31    &25 & 29 & 39 &36  &47  &57\\ \hline
\end{tabular}

\vl

Table 4.10. \label{SeffB} Predictions for $s(eff,B)$ assuming integer
values   for spin-spin and isospin-isospin  interaction parameters. The
errors induced in the prediction
 of masses  are below two per cent.

\subsubsection{Mesonic case}

For mesons the general splitting pattern is given by the following table.

\vl

\begin{tabular}{||l|l|l|l||}\hline\hline
meson&$\Delta s^I$ &$\Delta s^{spin}$&s\\ \hline
 $\pi$& $\frac{1}{4}K(0,M)$&$-\frac{3}{4}S(d,d,M)$&0\\ \hline
$\rho$& $\frac{1}{4}K(0,M)$&$\frac{1}{4}S(d,d,M)$&12\\ \hline
$\eta $ &$-\frac{3}{4}K(0,M) $&$-\frac{3}{4}S(d,d,M)$&6\\ \hline
$\omega $ &$-\frac{3}{4}K(0,M) $&$\frac{1}{4}S(d,d,M)$&12\\ \hline
$K^{\pm},K^0(CP=1)  $ &$\frac{1}{4}K(0,1,M)$& $-\frac{3}{4}S(d,s,M)$&5
\\ \hline
$K^0(CP=-1)$& $-\frac{3}{4}K(0,1,M)$ &$-\frac{3}{4}S(d,s,M)$&5\\ \hline
$K^{*,\pm},K^{*,0}(CP=1)  $ &$\frac{1}{4}K(0,1,M)$&
$\frac{1}{4}S(d,s,M)$&16
\\ \hline
$K^{*,0}(CP=-1)$& $-\frac{3}{4}K(0,1,M)$ &$\frac{1}{4}S(d,s,M)$&16
\\ \hline
$\eta'$ &$-\frac{3}{4}K(1,M)$&$-\frac{3}{4}S(s,s,M)$&19\\ \hline
$\Phi$ &$-\frac{3}{4}K(1,M)$&$\frac{1}{4}S(s,s,M)$&21\\ \hline
$\eta_c$ &$-\frac{3}{4}K(1,M)$&$-\frac{3}{4}S(c,c,M)$&184\\ \hline
$\Psi$ &$-\frac{3}{4}K(1,M)$&$\frac{1}{4}S(c,c,M)$&200\\ \hline
$D^{\pm},D^0(CP=1)$ &$\frac{1}{4}K(0,1,M)$&$-\frac{3}{4}S(d,c,M)$&72
\\ \hline
$D^0(CP=-1)$ &$-\frac{3}{4}K(0,1,M)$&$-\frac{3}{4}S(d,c,M)$&72\\ \hline
$D^{*,\pm},D^{*0}(CP=1)$ &$\frac{1}{4}K(0,1,M)$&$\frac{1}{4}S(d,c,M)$&83
\\ \hline
$D^{*0}(CP=-1)$ &$-\frac{3}{4}K(0,1,M)$&$\frac{1}{4}S(d,c,M)$&83
\\ \hline\hline
\end{tabular}

\vl

Table 4.11. \label{Isosmplit} Isospin and spin-spin splitting pattern for
mesons.

\vm

 The values of $S(M,i,j) $  and $K(M,i,j)$ can in principle
 deduced from the observed mass splittings. Complications however  result
from  the possible  mixing of $(I=0,J=0)$ and $(I=0,J=1)$ mesons.   \\
\\a) From $\rho-\pi$ mass splitting   one has $S(M,d,d)=12$.\\  b) If
$\omega$
  suffers no  mixing with $\Phi,\Psi,Ypsilon..$ the equality
$s(\omega)=s(\rho)$ implies
 $K(M,0,0)=0$ and $\Delta s_c(M)=-9$.
  \\ c)  $K^*-K$-splitting gives $S(M,d,s)= 16-5=11$ not far from
$S(M,d,s)=12$.\\ d) $D^*-D$ mass  difference $s(D^*)-s(D)= 83-72=11$
gives prediction $ S(M,d,c)=12$.  Note that again there is discrepancy of
one unit in the prediction of mass splitting.\\ e) Smallness of CP
breaking in $K-\bar{K}$ system implies $K(0,1)=0$. Same applies to
$D-\bar{D}$ and  $B-\bar{B}$ so that all nondiagonal elements of $K(i,j)$
must vanish, which in turn suggests that isospin-isospin interaction for
mesons vanishes identically in first order. \\ f) $F-D$ mass difference
gives $S(M,s,c)= \frac{4}{3}(s(s)-s(d)-1)$  and gives  $S(M,s,c)=16$ for
$s(s)=14$ with  predicted splitting too large by one unit and $S(s,c)=20$
for $s(s)=16$.

\vm

There are some discrepancies in the simplest picture without mixing
effects.\\ a)  $K-\pi$ mass splitting
 assuming $s(eff,s)=16$ is $s(K)-s(\pi)=7>5$.\\ b)   $\omega-\eta$
splitting
  $s(\omega)-s(\eta)=6$ gives,  assuming that $\eta$ does not
 contain $s\bar{s}$ pair,  $s(M,d,d)=8$,  which is smaller than    the
value suggested by $\rho-\pi$ splitting.  $\Phi-\eta'$ mass difference
 $s(\Phi)-s(\eta')= 21-19=2$ suggests $S(M,s,s)=4$ or $s(M,s,s)=0$.  Both
values    are  suspiciously small.\\
   The mixing of $\eta$ and $\eta'$  and $\eta_c$  provides  a nice
explanation
   of the discrepancies.  \\ c) The mass of $Ypsilon$ meson for
$s(eff,s)=16$ is predicted to be  $s(eff, Ypsilon)= 2s(b)-9+ \frac{1}{4}
(S(b,b)+K(2,2)) = 1868$, which gives $S(b,b)+K(2,2) =4\cdot 139=556$,
which is suspiciously large as compared with  the values associated with
lighter quarks.

\vm

One can consider  two manners to get rid of the $K-\pi$  discrepancy.\\
a)   The first  possibility is  the mixing of $\omega$ with its higher
mass  companions, which  allows  $K(M,0,0)=8$ and correct $K-\pi$ mass
difference. This however leads to negative mass squared for $\eta$:
$s(\eta)= -8$ in absence of mixings and  it therefore seems that
$K(M,0,0)\le 0$ is the only reasonable possibility. $K(M,0,0)<0$ in turn
implies $s(eff,\omega)>s(eff,\rho)$ and mixing effects can make the
situation only worse so that $K(0,0)=0$ is the only possibility and the
mixing of $J=1$ mesons  does not help.  Of course, $\Psi$ and  $Ypsilon$
$t\bar{t}$ $(J=1,I=0)$ mesons could mix  so that  one could avoid
anomalously large values of $s(M,b,b)$.  \\ b) The parameter $n_2(eff, d)$
fixing the mass of strange quark has negligibely  small effect on
 the values of CKM matrix elements  and therefore one could
 give up the  assumption that topological mixing matrices are identical
for mesons and baryons and assume that for mesonic quark one has

 \begin{eqnarray}
s(eff,s,M)&=&14
\end{eqnarray}

\noindent
  instead of $s(eff,s,B)=16$. This implies correct $K-\pi$ mass difference.
  This increases the value of  $s(eff,b,M)$ by $32$ units but the effect
is not large enough and one must still
 have $S(b,b)+= 4\cdot 75=300$ in order to get Ypsilon mass correctly.

\vm

If one assumes that $K(M,i,j)=0$ identically and no mixing for $J=1$
states one  obtains correct masses for $J=1$ mesons $\Phi,\Psi,B$  using
following values of
 parameters

\begin{eqnarray}
S(M,s,s)&=& 8\nonumber\\
S(M,c,c)&=&40\nonumber\\
S(M,b,b)&=& 300
\end{eqnarray}

\noindent   The predictions for the masses of $J=0$ partners are
systematically  too small

\begin{eqnarray}
s(eff,\eta)&=& 0<6\nonumber\\
s(eff,\eta')&=&13<19\nonumber\\
s(eff,\eta_c)&=& 159<185\nonumber\\
s(eff,\eta_b)&=& 1568
\end{eqnarray}

\noindent To get rid of the  remaining discrepancies one must assume mixing
 between
 neutral pseudo scalar mesons $\eta,\eta',\eta_c,\eta_b,\eta_t$.
 If mass squared is simply the quantum mechanical expectation for mass
squared
 of
 $\eta$ and  $\eta'$ and $\eta_c$ implies that  the masses of $\eta$ and
 $\eta'$ are actually smaller than the masses of their mixed partners  and
the effective value  of $s(M,i,j) $ derived from mass difference  is
expectation value, which contains also the  contribution of quark mass
differences.   In   the simplest scenario  mesonic isospin-isospin
interaction
 vanishes in first order for the lowest two generations at least,
$\eta$ mixes with $\eta'$ (pure $s\bar{s}$ pair) only  and  $\eta'$
mixes   with $\eta$ and $\eta_c$ only,....

\begin{eqnarray}
\eta_{phys}&=& c_1\eta+ s_1\eta'\nonumber\\
\eta'_{phys}&=& c_2(-s_1\eta+ c_1\eta') + s_2\eta_c\nonumber\\
\eta^c_{phys}&=& c_3(- s_2(-s_1\eta + c_1\eta') +
c_2\eta_c  ) + s_3\eta_b   \nonumber\\
{}.
{}.
{}.
\end{eqnarray}

\noindent   Various masses come out correctly if one has

\begin{eqnarray}
s_1^2&=&\frac{ s(\eta_{phys})}{s(\eta')}= \frac{6}{13}\simeq 0.462 \nonumber\\
s_2^2&=& \frac{ s(\eta'_{phys})-c_1^2s(\eta') }{s(\eta_c)-c_1^2s(\eta')}
=\frac{12 }{152}\simeq 0.079\nonumber\\
s_3^2&=& \frac{s(\eta^c_{phys})-c_2^2s(\eta^c)-c_1^2s_2^2s(\eta')}
{ s(\eta_b)-s(\eta_c)c_2^2-c_1^2s_2^2s(\eta') }\nonumber\\
&=& \frac{152\cdot 185-140\cdot 159-7\cdot 12}{152 \cdot 1568-1559\cdot 140
-7\cdot 12}
=\frac{5776}{216426}\simeq .0267\nonumber\\
{}.
{}.
 \end{eqnarray}

\vm

 To sum up, the discussion  leads to a definite predictions for
$\eta,\eta',\eta_c$ mixing
   and fixes
the values  of the parameters $S(i,j)$ and $K(i,j)$

\vl

\begin{tabular}{||l|l|l|l|l|l|l||}\hline\hline
S(d,d)&S(d,s)& S(d,c)&S(s,s)&S(s,c)&S(c,c)&S(b,b) \\ \hline
12&     12&     12    &  8   &16 &     40&    300\\ \hline\hline
\end{tabular}

\vl

Table 4.12. \label{Mesosplitparam} The values of spin-spin interaction
parameters for mesons.  Isospin-isospin interaction is assumed to vanish
in first order for mesons.

\vm

The first order contributions to the masses of 'diagonal' mesons are
reproduced correctly.  Situation is different for 'non-diagonal' mesons.
$K^*-K$ and  $D^*-D$ mass  splittings  are  predicted to be too large by
one unit and there are large errors in the
 predictions for  $D$,$ F$  and $B$ meson masses, which will be discussed
separately.

\vm

\begin{tabular}{||l |l |l |l |l |l|l |l|l|l|l|l| |}\hline\hline
meson&$\pi$&$\rho$&$\eta$&$\omega $&$K$&$K^*$&$\eta'$&$\Phi$ &$\eta_c$&
$\Psi$&
 Ypsilon\\ \hline
$s(eff,M,exp)$&0&12 &6&12&5&16&19&21&185&200&1868 \\ \hline
$s(eff,M,pred)$&0&12 &6&12&5&17& 19&21&185&200&1868 \\ \hline\hline
\end{tabular}

\vl

Table 4.13. Predictions for $s(eff)$ for mesons  assuming integer values  for
spin-spin and isospin-isospin  interaction parameters. $D,F$ and $B$
mesons are not included in the table.

\vm

\noindent   The values of parameters allow to deduce the parameter $\Delta
s_c $  characterizing the strength  of color Coulombic interaction  in
order $O(p)$. The results are

\begin{eqnarray}
\Delta s_c(B)&=&3c(B)=-3+\frac{1}{2}\nonumber\\
\Delta s_c(M)&=& c(M)=-9
\end{eqnarray}

\noindent  These parameters fix the the masses of proton and pion and it
would
 be nice if one could really predict the values of these parameters. A QCD
inspired estimate for
 the color binding energies

\begin{eqnarray}
\Delta m_c(B)&=& -\frac{3C}{2}\frac{\alpha_s (r_B)}{r_B}\nonumber\\
 \Delta m_c(M)&=& -\frac{C}{2}\frac{\alpha_s(r_M)}{r_M}
\end{eqnarray}

\noindent  where $C(3)=8/3$ is the value of Casimir operator for triplet
representation,  is consistent with TGD estimate provided the value of
color coupling strenght in
 mesonic length scale $r_M$ larger than baryonic length scale is much
larger than in baryonic
 length scale $r_B$. For proton the estimate $3/18=\Delta m^2/m^2=2\Delta
m_c/m$ gives estimate
 $\alpha_s\sim  r_{B}m_p/16\sim 0.06$, which is reasonably near to
$\alpha_s \sim 0.12$. For pion the corresponding estimate $\alpha_s \sim
3m_pr_M/(4\sqrt{2})$, which
 is of order one.

\subsection{Condensate level mixing}

The following table shows the predictions for the parameter $s$
determining  mass
 squared  in order $O(p)$ for   heavy hadrons containing charmed and
bottom quarks. Topological mixing and  color magnetic hyperfine and
isospin-isospin splittings  (which are rather small effects)  are taken
into account, when corresponding parameters
 are known.

\vl

\begin{tabular}{||l|l|l||}\hline \hline
hadron &$s$& $s_{exp}$\\ \hline
$\Lambda_c$&$s(\lambda) +s(c)-s(s)=108$ &108\\ \hline
$\Lambda_b$&$s(\lambda) +s(b)-s(s)=878 $       &614  \\ \hline
$D$&$s(c)-\frac{3}{4}S(d,c)=90$    &72  \\ \hline
$B$&$s(b)-\frac{3}{4}S(d,b)=869-32\Delta-\frac{3}{4}S(d,b)$     &588  \\ \hline
\end{tabular}

\vl

Table 4.14. \label{Anomal}  The predictions for certain hadron masses
 compared with experimental masses for  $s(s)=16-\Delta$ ($\Delta=2$ is
suggested by
 $K-\pi$ mass difference.

\vm

\noindent From the table it is clear the values of mass squared  of
 hadrons ($B$ and $\Lambda_b$) containing one bottom quark are much lower
than the predictions following from the proposed scenario. Hyper fine
splittigs and isospin-isospin interactions certainly cannot help in the
problem.  For $D$  meson containg one c quark ( hyper fine splitting has
been taken into
 account) ,  the effect seems to be   present but is considerably smaller.
 For $\Lambda_c$ the  first order prediction  is exact.

\vm

  A possible explanation is that $b$ quark  spends part of its time  on
$k=107$ level rather than $k=103$ level so that average mass squared is
reduced.   Since  the  prediction for the mass of Ypsilon is essentially
correct  it seems that inside  Ypsilon the mechanism is not at work.   The
only manner to escape contradiction is to assume that $b$ quark condenses
on $u,d$ or $s$ quark, whose primary condensate level is also  $k=107$,
which in turn condenses on hadronic level, which corresponds to $p$ near
$M_{107}$ and has also $k=107$.     Inside Ypsilon there is no place for
$b$ to condense and  therefore Ypsilon remains heavy.  The same mechanism
applies to $\Psi$.  If   quark prefers  to condense on  antiquark rather
than quark then one could understand why the effect is not  seen for
$\lambda_c$.  One must however remember that the effect is small for  c
quark and some other explanation might work equally well.

\vm

Condensate level mixing can be described quantum mechanically in same
manner as  the topological mixing. Physical state at $k=107$ is
superposition of states,   where the primary condensation level of $b$
($c$) quark is  either $k=103$ or $k=107$ and mass squared is quantum
mechanical expectation value   for masses for these two primary
condensation levels

\begin{eqnarray}
\vert b_{phys}\rangle &=& cos(\phi)\vert  b_{103}\rangle +
sin(\phi)\vert  b_{107}\rangle \nonumber\\
s_{phys}(b)&=&  cos^2(\phi) s(b,k=103)+ sin^2(\phi)n(b,k=107)\nonumber\\
s(b,103)&=& 5+ 2^4( 54+\Delta)\nonumber\\
s(b,107)&=& 5+54+\Delta\nonumber\\
s(s)&=& 16-\Delta
\end{eqnarray}

\noindent This equation  has obvious generalization to the case of $c$
quark with $s(c)= 3+ 6\cdot 2^4$   and one obtains the following mass
formula

\begin{eqnarray}
s_{phys}(b)&=& 59+\Delta 15\cdot (54+\Delta) cos^2(\phi_b)\nonumber\\
 s_{phys}(c)&=&  9+ 15\cdot 6 cos^2(\phi_c)\nonumber\\
cos^2(\phi_b)&=& \frac{n_b}{ 15\cdot (54+\Delta)  }\nonumber\\
cos^2(\phi_c)&=& \frac{n_c}{ 90  }
\end{eqnarray}

\noindent  No Planck mass requirement implies the expressions  for  mixing
 angle.

\vm

 The requirement that $B$ and $D$ masses are predicted
correctly implies

\begin{eqnarray}
n_b&=& 529+\Delta + \frac{3}{4}S(M,d,b)\nonumber\\
n_c&=&  72\nonumber\\
sin^2(\phi_b)&=& \frac{281+14\Delta-\frac{3}{4}S(d,b)}{ 810+15\Delta }
  \nonumber\\
sin^2(\phi_c)&=&\frac{1}{5}
\end{eqnarray}

\noindent where $\Delta s^I (D/B)=0$ (no CP breaking in first order)
and $\Delta s^{spin}(D)=-9$ are used.

\vm

In the baryonic case one can  derive constraints on  the values of the
mixing  angles
 from $\Lambda_b$ and $\Lambda_c$ masses. For c quark baryonic mixing
angle  vanishes.
  A possible explanation is that condensation of c quark to anti quark  is
more probable process than condensation on quark.
 The  requirement that baryonic mixing angle  for $b$ quark is identical
with the  mesonic one
 fixes the value of the mesonic spin-spin interaction parameter
$S(M,d,b)$,
 the value of the mixing angle and the value of $s(eff,mix,b)$

\begin{eqnarray}
S(M,d,b)&=& \frac{4}{3}(s(\Lambda_b)-s(B)+s(s)-s(\Lambda))\rightarrow 32
\nonumber\\
sin^2(\phi_b)&=& \frac{57}{168}\simeq 0.34\nonumber\\
s(eff,mix,b)&=&s(B)+\frac{3}{4}S(M,d,b)= 612
\end{eqnarray}

\noindent   The result means that b quark spends about one third of its
time   on  lower condensate level.

\subsection{Summary: hadronic mass formula in first  order }

For the convenience of the reader the   hadronic mass formula  in first
order is  summarized. The value of the coefficient  $s(H)$ in $M^2(H)=
s(H)p+...$  giving leading contribution to
 the hadron mass (pion being exception) is

\begin{eqnarray}
s(H)&=&\sum_i s(q_i,eff)+ \sum_{pairs}S(i,j) \bar{s}_i\cdot \bar{s}_j +
\sum_{pairs}K(g(i),g(j))\bar{I}_i\cdot \bar{ I}_j+\Delta s_c
\end{eqnarray}

\noindent a)  The values of $ s(q_i,eff)$ correspond to quark masses,  when
topological mixing effects are taken into account. For hadrons containing
both  charmed or b  quarks and u,d,s quarks  there is also the mixing of
primary condensation
 levels present and different value of $s(q_i,eff)$ must be used.  $K-\pi$
mass difference is too large by  2 units unless one uses $s(M,s)=14$
instead of the baryonic value $s(B,s)=16$:
 CKM matrices are essentially identical in both cases.

\vl

\begin{tabular}{|l|l|l|l|l|l|l||}\hline\hline
baryonic  quark&d&s&b&u&c&t\\ \cline{1-7}\hline
$s(eff)$&9&16&869&9&99&$ 3+57\cdot 2^{10}$\\ \hline
$s(eff,mix)$&9 &16 &$612$& 9&99&?\\ \hline\hline
mesonic  quark&d&s&b&u&c&t\\ \hline
$s(eff)$&9&14&901&9&99&$ 3+57\cdot 2^{10}$\\ \hline
$s(eff,mix)$&9 &14 &612& 9&81&?\\ \hline\hline
\end{tabular}

\vl

Table 4.15. \label{Seffquark} The known values of $s(eff,q_i)$ in the
scenario
 providing a solution for the  spin crisis of proton. The second row
contains  the effect  of condensate level mixing in hadrons containing
$u,d$ or $s$ quark and $c$ quark or $b$ quark.  For baryons the  effect is
not present for c quark and the value given in
 parenthesis must be used. For mesons $s(s)=16$ would predict $K-\pi$
mass  squared difference two  units too large. For top quark  $k=97$
condensation level is assumed in the mass formula.

\vm

\noindent b)  The elements of the integer matrices $S(i,j)$ and
  $K(g(i),g(j))$ parametrize  color magnetic spin-ispin interaction
 and  isospin-isospin interaction and are given in the tables below.

\vl

\begin{tabular}{||l|l|l|l|l|l|l|l||}\hline \hline
$K(0,0)$&$K(0,1)$&$K(1,1)$&$S(d,d)$&$S(d,s)$&$S(s,s)$&$\Delta s_c$\\ \hline
$ 4 $&$0 $&$ 10 $&$5-\frac{1}{3}$ &$ 12+\frac{2}{3}$& $7-\frac{1}{3}$&
$-1+\frac{1}{6}$
\\ \hline
$ 4 $&$0 $&$ 10 $&$4$ &$ 12$& $6$&$-1$\\ \hline\hline
\end{tabular}

\vl

Table 4.16. \label{Isobparam}   Estimates for the  parameters describing
spin-spin
 and isospin-isospin interactions  for baryons. Fractional valued
parameters reproduce low lying hadron masses exactly and integer valued
parameters   reproduce masses with  errors not larger than $\vert \Delta s
\vert =1$.

\vl

\begin{tabular}{||l|l|l|l|l|l|l|l||}\hline\hline
S(d,d)&S(d,s)& S(d,c)&S(s,s)&S(s,c)&S(c,c)&S(d,b)&S(b,b)\\ \hline
12&     12&     12&      8 & 16     &40     &32 & 300\\ \hline\hline
\end{tabular}

\vl

Table 4.17. \label{Isomparam}   Estimates for the  parameters describing
spin-spin  interactions  for mesons assuming  $s(eff,ss)=14$ for mesonic
strange quark. Isospin-isospin interaction vanishes for mesons in lowest
order.

\vm

\noindent c)  The  values of the integer $\Delta s_c$  parametrizing
color Coulombic binding energy for baryons and mesons are

\begin{eqnarray}
\Delta s_c(B)&=&-3+\frac{1}{2}\nonumber\\
\Delta s_c(M)&=& -9
\end{eqnarray}

\noindent  If the parameters are assumed to be integers one has
 $\Delta s_c(B)=-3$.\\ d)  For neutral pseudoscalar mesons $\eta,\eta',
\eta_c$ also mixing effects are important  in first order and the values
and are obtained by calculating  the masses of $\eta,\eta',\eta_c$ using
previous formulas and using following formulas to
 take mixing effects into account

\begin{eqnarray}
s(\eta_{phys})&=& s_1^2s(\eta')\nonumber\\
s(\eta'_{phys})&=&  c_1^2c_2^2s(\eta') + s_2^2s(\eta_c)\nonumber\\
s(\eta^c_{phys})&=& c_1^2s_2^2c_3^2s(\eta') +
c_1^2c_2^2c_3^2s(\eta_c)+s_3^2s(\eta_b)\nonumber\\
 s_1^2&=& \frac{6}{13}\nonumber\\
s_2^2&=& \frac{12}{52}\nonumber\\
s_3^2&=& \frac{5776}{216426}
 \end{eqnarray}

\vm

In baryonic case first order contributions are reproduced  exactly for
 fractional values of $S(i,j),K(i,j)$ and for integer values errors are
not larger than $\vert \Delta s \vert =1$.
 If $s(s,M)=14$ is assumed the errors in mesonic sector are associated
with  spin-spin splittings of nondiagonal
 mesons $K-K^*,D-D^*$,   which are predicted to be too large
 by one unit. For $s(s,M)=16$   the value of $s(K)$ is predicted to be two
units too large.  One cannot exclude
 the possibility of some unidentified contribution to the masses of
nondiagonal  mesons.

\subsection{Second order contribution to mass squared and isospin
  splittings }

Second order  terms  in cm contribution to the  masses of the quarks
imply definite  isospin splitting pattern for hadron masses.  The
splittings are not in accordance with experimental splittings as the last
columns of the table below show. Proton-neutron mass squared  difference
is of correct sign but too large by a factor $16$.
 For $DDD$  ($\Delta^ -$)  configuration $X$ vanishes whereas
experimentally the mass in general increases with decreasing charge.
There are several sources  to second order term in mass squared.

\vm

\noindent a) Coulombic splitting:    the general form of Coulombic mass
splitting in second order is given by

\begin{eqnarray}
\Delta X(Coul)&=& \sum_{pairs} Q_kQ_l  D_e(k,l)
\end{eqnarray}

\noindent b) Electromagnetic spin-spin interaction.  The general form of
 the magnetic interaction
 is

\begin{eqnarray}
\Delta X(magn)&=& \sum_{pairs} Q_kQ_l\bar{S}_k\cdot \bar{S}_l D_m(k,l)
\end{eqnarray}

\noindent The strongest assumption is that the parameters $D_e(k,l)$ and
 $D_m(k,l)$ are same for all quark pairs and
 all baryons and mesons respectively.  A more realistic assumption is that
$D_e$ are different for mesons and baryons. Still one step towards realism
is to notice that Coulombic and magnetic energies are inversely
proportional to  the first and  third power of electromagnetic radius of
hadron respectively so  that  an appropriate  hadron dependent scaling
factor could be present in the parameter.   $D_m(k,l)$ could also  depend
on quark pair (being inversely proportional to the product of quark masses
in quark model). An interesting possibility is that $D_m(k,l)$ is
proportional to the matrix  $S(k,l)$ associated with color hyper fine
interaction. This could be the case  since the matrix in question
summarizes information about geometry (average quark-quark distance)  and
quark masses. Therefore an interesting hypothesis to be tested is

\begin{eqnarray}
 D_m(k,l)&=& d S(k,l)
\end{eqnarray}

\noindent  There is considerable information on elements of $S(k,l)$ at
use.
  In the following it is useful to take  the notational conventions

\begin{eqnarray}
C_{m}k,l)&=&\frac{D_m(k,l)}{36}\nonumber\\
C_e(k,l)&=& \frac{D_e(k,l)}{9}
\end{eqnarray}

\noindent at use.

\vm

There are also interactions, which produce constant shifts depending on
isospin
 and spin multiplet, which are important in p-adic regime since the
addition of small  power of two to second order term may change the real
counterpart of term radically.  \\  a) Second order contribution to color
spin-spin interaction. This interaction  introduces only constant shift
between $J=3/2$ and $J=1/2$ baryons and between $J=1$ and $J=0$
 mesons and the form of this contribution is same as in $O(p)$  case. \\
b) Color Coulombic interaction. This introduces only constant shift for
the  multiplets.  Similar shift results from the sea contribution to mass
squared, which is  automatically of order $O(p^2)$. \\ c) Isospin-isospin
interaction. If the interaction is present for the fermions of same
generation the interaction
 produces characteristic splittings between different isospin multiplets.
\\ At this stage the best manner to treat  these terms is to allow a
constant
 shift depending on electroweak and spin multiplet.

\vm

 The assumption that isospin splittigns result from $O(p^2)$ effects has
important consequence: the value of $s(H)$ is same for all hadrons inside
isospin multiplet.   This requirement allows one to derive stringent
bounds on the value of the fundamental mass scale $m_0^2$.
 Electron mass fixes in principle the value of $m_0^2$ but
 second order corrections to electron mass (topological mixing, Coulombic
energy) induce
 changes in mass scale and
  for heavier hadrons small change can induce even a change in the value of
 $s$ and the values of $X$ suffer shift (change is proportional to $\delta
M^2$, where  $\delta$ is the change in basic mass scale).  A possible
manner to get rid of inaccuracy is to deduce the
 scale factor by comparing two  known  hadrons known to have identical
second order correction to mass squared
 (hadrons of $M_{89}$ hadron physics are  ideal in this respect!).  The
fit of masses based on  electron mass formula without Coulombic and mixing
corrections has some  flaws:  for instance,
 the values of $s$ differ by one unit for $\Xi^-$ and $\Xi^0$: this is a
clear indication about the presence of corrections to electron mass.    In
order to get consistency one must increase or decrease basic  mass scale a
little bit.
 Since corrections to
electron mass seems to  be positive the decrease of  mass scale comes in
question so that $s$ and $X$   for heavier baryons tend to increase. The
minimal change in scale corresponds to

\begin{eqnarray}
m_0^2&\rightarrow & m_0^2\frac{5+2/3}{5+2/3+ \delta}\nonumber\\
\delta &\simeq& 0.0515
\end{eqnarray}

\noindent  An attractive posibility is that $\delta $ super position of a
small
 number of  powers of two: simplest alternative is $\delta =1/64\simeq
0.0516$. With this assumption  $\Delta^{-}$ has  still  $s$ larger by one
unit than its companions  and has $X(\Delta^-)=1/64$: the experimental
uncertainty in $\Delta^--\Delta^{++}$  splitting is  much  larger than the
reduction in  $\Delta^- $ mass needed to get rid of the difficulty.
 The proposed  value of $\delta$ leads to an essentially  correct
prediction for $Z^0$ mass  and $W$ mass is predicted correctly taking
Coulombic corrections into account.

\vm

The reliable  determination of the parameters $C_e,C_m$ and shifts
$\Delta (B)$ depending on multiplet is not a trivial task and there are
some delicate points involved. \\ a)  The real counterpart  of the  second
order term is very sensitive to the changes in
 its p-adic counterpart:  relatively  small fractional contribution, say
constant shift,
 could change totally the splitting scenario unless it corresponds to
sufficiently large negative  power of $2$. The reason
 is that  the p-adic power  expansion of fractional numbers has as its
lowest
 term large p-adic number of order $p$.   Two large terms of this kind
just below $p=M^{107}$ can
 hower sum up to number which is essentially zero modulo p.\\ b)   The
appearence of $1/60$ factors in the cm contribution to second order  term
suggests that natural unit for second order term is $1/60$. This is not
the case.  The point  is that $1/60$ is not expressible as sum of finite
number of powers of two and this in turn implies
 the existence of several values of $X_{eff}$ with the property that
$pX_{eff}$ maps to same  $X_{eff}$ in canonical correspondence.  For
instance, for $n= 8$ p-adic numbers $8p/120$,$36p/120$ and $ 68p/120$  map
 to $n/60$ for $p=M_{107}$.
 A natural unit for second order term  is $1/64$ since it corresponds to
the natural unit of mass squared for the  representations of Super
Virasoro with broken conformal symmetry. The second nice feature is that
the p-adic counterpart of $X/64$ for  $X<64$ is just $X/64$ since the
inverse of $64$ is $2^{101}$ in good approximation.
 \\ c) The calculation of first order terms for p-adic expansions of
fractional number $m/n$  reduces to that of finding the inverse of the
denominator $n$ in the finite field
 $G(p=M_{107},1)$ consisting of integers smaller than $p$. The inverses of
following numbers
 will be needed often:

\begin{eqnarray}
\frac{1}{3}&=& 2^{106}+2^{104}+....\nonumber\\
\frac{1}{5}&=& 2^{105}-2^{103}+..... \nonumber\\
\frac{1}{25}&=& 2^{103}-2^{102}+2^{100}-2^{99}-2^{98}+....\nonumber\\
\frac{1}{15}&=& 2^{103}+2^{99}+....\nonumber\\
\frac{1}{120}&=& 2^{100}+2^{96}+...
 \end{eqnarray}

\noindent Using  these one can  easily generate the real counterparts of
the  second order term in mass squared.

\subsubsection{ Baryonic case}

Consider first the determination of the parameters
$C_m(d,d)=C_m,C_e(d,d)=C_e$  and $\Delta X_0$ for nucleons and $\Delta$
resonances.
 The following table summarizes various electromagnetic contributions to
the coefficient o
 second order term for nonstrange  baryons.   The coefficient of various
contributions are normalized to integers in order to facilitate
calculations: one has  $E_m\equiv 3C_m$ and
 $C_e=D_e/9$.
 Listed are also the p-adic counterparts of $X_{eff}$ in order to
facilitate  calculations.

\vl

\begin{tabular}{||l|l|l|l|l||}\hline\hline
baryon &$\frac{\Delta X(magn)}{E_m}$ &$\frac{\Delta X(coul)}{C_e}$&X
&$64X_{eff}$\\
\hline \hline p&2 &0 &  $\frac{13}{30}$&  25
 \\ \hline\hline
n&$2$ &$-1$  &    $\frac{18}{30 }$& 28
 \\ \hline\hline
  $\Delta^{++}$ &$4$ &$4$ &$\frac{24}{30}  $ & 41   \\ \hline\hline
 $\Delta^+$ &$0$ &0  &    $\frac{13}{30}$ &54
  \\ \hline\hline
 $\Delta^0$& $-1$&$-1$ & $ \frac{18}{30 }$&50 \\   \hline\hline
 $\Delta^-$& $1$  &$1$ &  $ 0$  & $\leq 63,\ge 41$
\\ \hline\hline
 \end{tabular}
\vl

Table 4.18. \label{Barisospinsplit} Various contributions to iso-spin
splitting
 for nonstrange baryons. \vm

The masses are taken from the most recent Particle Data Tables and .  the
values of $X_{eff}$  are   subject  to a shift of $\Delta X=\pm 2$
resulting from the error in average mass.
  The relative error for  $\Delta^{-}-\Delta^{++}$ mass difference in
older Particle Data Tables  and is almost as large as the mass
difference.  In the most recent Particle Data Table this mass difference
is not
 reported.  The use of the rather large average value of
$\Delta^{-}-\Delta^{++}$ mass difference
 would give $s(\Delta^-)=s(\Delta^{++})+1$, which cannot hold true if
isospin splittings are second order effect.
 Fortunately, only a slight
 change in $\Delta^-$ mass saves the situation.

\vm

\noindent  a) Proton and neutron
masses give the conditions

\begin{eqnarray}
C_e&=& X_{eff}(p)-X_{eff}(n) +\frac{5}{30}= -\frac{3}{64}+\frac{5}{30}
\nonumber\\
 \Delta (N) &= &-2E_m+ X_{eff}(p)-\frac{13}{30}=-2E_m+ \frac{25}{64}-
\frac{13}{30}
\end{eqnarray}

\noindent b)  One can solve the values of the parameters $E_m,\Delta
(\Delta)$  from $\Delta$
 masses and obtains two $\Delta$  masses as predictions. It is useful to
express
 the solution in terms
 of  $\Delta^0-\Delta^+$ mass difference $\delta$.

\begin{eqnarray}
E_m&=&\frac{3}{64}+ \delta= \frac{7}{64} \nonumber\\
\Delta (\Delta)&=& X(\Delta^+)-\frac{13}{30} \nonumber\\
X(\Delta^{++})&=& X(\Delta^+) +\frac{1}{30} -4\delta=
\frac{41}{64}+\frac{1}{30}
\nonumber\\
X(\Delta^-)&=& X(\Delta^+) -\frac{2}{15}+\delta =
\frac{56}{64}-\frac{2}{15}
\nonumber\\
\delta&=&X(\Delta^+) -X(\Delta^0)= \frac{4}{64}
\end{eqnarray}

\noindent  From the formulas for the masses it is clear that masses are
very
 sensitive to the value of the mass difference $\delta$.\\
 d) The masses of $\Delta^{++}$ and  $\Delta^{-}$ come out as follows:

\vl

\begin{tabular}{||c|c|c|c||}\hline\hline
baryon &X&$(pX_{tot})_R$& $X_{eff}$\\ \hline \hline
 $\Delta^{++}$&$   \frac{38}{64}+\frac{1}{30}  $ &$\frac{40}{64}+
\frac{1}{15\cdot
32}$&$\frac{41}{64} $
 \\ \hline
$\Delta^+$& $ \frac{54}{64}    $  &$\frac{54}{64}$&$\frac{54}{64}$\\
\hline\hline $\Delta^0$& $ \frac{50}{64}    $
&$\frac{50}{64}$&$\frac{50}{64}$ \\ \hline\hline $\Delta^-$& $
\frac{58}{64}-\frac{2}{15}    $  &$\frac{50}{64}-\frac{1}{120}$ &$?$ \\
\hline\hline
 \end{tabular}

\vl

Table 4.19. \label{Deltaisospinsplit} The predictions for  the isospin
 splittings of
$\Delta$ resonances.

\vm

$\Delta^{++}$ mass is predicted  correctly within experimental
uncertainties.  The masses  of $\Delta^-$ and $\Delta^0$ are predicted  to
be identical with  good accuracy: the prediction is within the error bars
of the earlier  Particle Data Table data on the
 $\Delta^--\Delta^{++}$ mass difference.
  Unfortunately Particle Data tables give no information about this mass
difference.

\vm

Consider next strange baryons, for which second order contributions to
mass are
 listed in  the table below.

\vl

\begin{tabular}{||l|l|l|l|l||}\hline\hline
baryon &$\Delta X(magn)$ &$\Delta X(coul)$&X
&$64X_{eff}$\\ \hline \hline
$\lambda$& $6C_m(d,s)+6\Delta C_0$  &$-C_e(d,s)-2\Delta C_e^0$ &
  $\frac{18}{30 }$
&$63$\\ \hline\hline
$\Sigma^+$& $ 6C_m(d,s)-12\Delta C_0   $  &$4\Delta C_e^0$ &
 $\frac{13}{30}$  &
39 \\ \hline\hline
 $\Sigma^0$& $   6C_m(d,s)+6\Delta C_0$  &$-C_e(d,s)-2\Delta C_e^0$ &
 $\frac{18}{30}$&
45\\ \hline\hline
 $\Sigma^-$& $     - 3C_m(d,s)-3\Delta C_0$  &$C_e(d,s)+  \Delta C_e^0$ &
  $
0$   &60\\ \hline\hline
$\Sigma^{*+}$& $4\Delta C_0$  &$4\Delta C_e^0$ & $\frac{13}{30}$ &  4 \\
\hline\hline
 $\Sigma^{*0}$& $-3C_m(d,s)-8\Delta C_0$  &$-C_e(d,s)-2\Delta C_e^0$ &
 $\frac{18}{30}$
&4\\ \hline\hline
 $\Sigma^{*-}$& $3C_m(d,s)+\Delta C_0$  &$C_e(d,s)+  \Delta C_e^0$ &  $
0$  &4\\ \hline\hline
$\Xi^0$& $6C_m(d,s)+6\Delta C_1$  &$-C_e(d,s)-2\Delta C_e^1$ &
 $\frac{18}{30
}$&7     \\ \hline\hline
 $\Xi^-$&$-3C_m(d,s)-3\Delta C_1$   &$C_e(d,s)+\Delta C_e^1$ &  $0 $
&29\\
 \hline\hline
 $\Xi^{*0}$& $-3C_m(d,s)-8\Delta C_1$  &$-C_e(d,s)-2\Delta C_e^1$ &
$\frac{18}{30 }$&0   \\
\hline\hline
  $\Xi^{*-}$& $3C_m(d,s)+\Delta C_1$  &$C_e(d,s)+\Delta C_e^1$ &  $0 $
&13\\
 \hline\hline
 $\Omega^-$& $-3C_m(s,s)$  &  $ C_e(s,s)$ &$0$  &25\\ \hline\hline
\end{tabular}

\vl

Table 4.20.\label{Strangeisospinsplit}  Various contributions to isospin
splitting  for strange baryons assuming that magnetic and Coulombic
interaction term is
 different for different quark pairs.  The notations $\Delta C_0=
C_m(d,d)-C_m(d,s)$,
 $\Delta C_1= C_m(s,s)-C_m(d,s)$, $\Delta C_e^0= C_e(d,d)-C_e(d,s)$ and
$\Delta C_e^1=  C_e(s,s)-C_e(d,s)$  are used.

\vm

Consider first $\Sigma$ and $\Sigma^*$ baryons.  The isospin splittings of
 $\Sigma^*$ are poorly known and one must use only $\Sigma$:s to deduce
the values of $C_m(d,s), C_e(d,s)$
 and multiplet shift $\Delta (\Sigma)$ in terms of the known parameters.
Some amount  of  linear  algebra gives

\begin{eqnarray}
\Delta (\Sigma)&=&  X_{eff}(\Sigma^-) +3C_m(d,d)-C_e(d,d)=\frac{3}{32}-
\frac{1}{6}\nonumber\\
C_e(d,s)&=& -\Delta (\Sigma)+ X(\Sigma^0)-6C_m(d,d)+2C_e(d,d)-
\frac{18}{30}=
\frac{19}{64}- \frac{1}{10}\nonumber\\  C_m(d,s)&=& \frac{\frac{13}{30}-
 X_{eff}(\Sigma^+)+ \Delta
(\Sigma) } {4}+C_m(d,d)+ C_e(d,d)-C_e(d,s)\nonumber\\
&=&-\frac{7}{16}+\frac{1}{3}    +\frac{1}{3 \cdot 256}\simeq -\frac{1}{6}
\nonumber\\
\
\end{eqnarray}

\noindent  (recall the definitions  $E_m=3C_m(d,d)$ and $C_e(d,d)=C_e$).
 The knowledge of $\Sigma^*$ masses would allow the deduction of $\Delta
(\Sigma^*)$ and
 prediction of splittings inside $\Sigma^*$  multiplet. $C_m(d,s)\simeq
-\frac{1}{6}$ has  much larger  absolute value than  $C_m(d,d)\simeq
\frac{1}{32}$  but has opposite sign whereas   $C_e(d,s)
=\frac{19}{64}-\frac{1}{10}$  is of same  order of magnitude as
$C_e(d,d)\simeq \frac{1}{6}$.

\vm

The splittings in  $\Xi^*$ and $\Xi$ multiplets are  better known and one
can
 deduce the  values of the parameters $C_m(s,s),C_e(s,s),\Delta (\Xi)$ and
 $\Delta (\Xi^*)$ but no predictions are posible. After some algebra one
obtains

\begin{eqnarray}
\Delta (\Xi)&=& \frac{  X_{eff}(\Xi^{0})+2X_{eff}(\Xi^{-})-\frac{18}{30}-
C_e(d,s) }{3}
= \frac{15}{64}- \frac{1}{6}+\frac{1}{3\cdot 64}\simeq -\frac{5}{12}
\nonumber\\
C_m (s,s)&=&
\frac{X_{eff}(\Xi^{*0})-X_{eff}(\Xi^{0})-X_{eff}(\Xi^{*-})+
X_{eff}(\Xi^{-})+25C_m(d,s)     }{5} \nonumber\\
&\simeq&
-\frac{5}{32}+\frac{13}{60\cdot 64}\nonumber\\
 C_e(s,s)&=& 3C_m(s,s)+X_{eff}(\Xi^{*-}) -\Delta (\Xi)=
\frac{1}{2}+\frac{1}{6}-\frac{19}{60\cdot 64}+\simeq \frac{2}{3}
\nonumber\\
\Delta (\Xi^*)&=& \Delta (\Xi)+ 5C_m(d,s) +4C_m(s,s) +X_{eff}(\Xi^{*0})-
X_{eff}(\Xi^{0})
\nonumber\\
&\simeq & -\frac{17}{64}-\frac{1}{6}+\frac{1}{20\cdot 64}\simeq -
\frac{5}{12}
\nonumber\\
\
 \end{eqnarray}

\noindent   Notice that $C_m(s,s)$ does not differ numerically very  much
from  $C_m(d,s)$.  In case of $\Lambda$ and $\Omega$ one multiplet shifts
$\Delta (\lambda)$ and  $\Delta (\Omega)$ are the only unkown quantities
and can be derived from the known values of $X_{eff}$.   Rather
frustratingly, no actual tests for the scenario are obtained due to the
poor knowledge of $\Sigma^*$ splittings.

\subsubsection{ Mesonic case}

The following table represents contributions to mesonic second order term
in  mass squared and also the a values of $X_{eff}$ deduced from mass fit.

\vm

\begin{tabular}{||l|l|l|l|l|l|l||}\hline\hline
meson &$\frac{\Delta X(magn)}{E_m}$ &$\frac{\Delta X(coul)}{C_e}$&X&
$\Delta X^I/\Delta^I_0$&$\Delta
X^s/\Delta^{s}$ &$64X_{eff}$\\ \hline \hline
$\pi^0$&$\frac{5}{4}$ &$-\frac{5}{2}$  &    $\frac{13}{30 }$&
$-\frac{1}{4}$& $-\frac{3}{4}$  &      24\\  \hline\hline
$\pi^+$&-1 &2&  $\frac{13}{30}$&    $-\frac{1}{4}$& $-\frac{3}{4}$  &26 \\
\hline\hline
 $\rho^0$&$-\frac{5}{12}$ &$-\frac{5}{2}$  &    $\frac{13}{30 }$&
$  -\frac{1}{4}$& $\frac{1}{4}           $&29 \\  \hline\hline
$\rho^+$&$\frac{2}{6}$ &$2$  &    $\frac{13}{30 }$&
$ -\frac{1}{4}$& $\frac{1}{4} $&24\\  \hline\hline
$\eta$& $  \frac{5}{4}$&$-\frac{5}{2}$ & $ \frac{13}{30 }$&
$ -\frac{1}{4}$& $-\frac{3}{4} $
&19\\   \hline\hline
 $\omega$& $ - \frac{5}{12}$&$-\frac{5}{2}$ & $ \frac{13}{30 }$&
$ \frac{3}{4}$& $\frac{1}{4}  $
&51 \\   \hline\hline
$K^+$ &$-1$ &$2$ & $\frac{13}{30}$  & $ 0$&  $ -\frac{3}{4} $  &6\\
\hline\hline
 $K^0$ &$\frac{1}{2}$ &-1  &    $-\frac{10}{30}$ &   $0$&$-\frac{3}{4}  $
 &11     \\
\hline\hline
 $K^{*+}$ &$\frac{1}{3}$ &$2$ & $\frac{13}{30}$  & $0 $&$  \frac{1}{4}  $
  &39\\
\hline\hline
 $K^{*0}$ &$-\frac{1}{6}$ &-1  &    $-\frac{10}{30}$ &   $0$
&$ \frac{1}{4} $
&49      \\ \hline\hline
$\eta' $& $\frac{1}{2}$  &$-1$ &  $-\frac{10}{30} $  &$ \frac{3}{4}$&
 $-\frac{3}{4} $
&10 \\ \hline\hline
$\Phi $& $-\frac{1}{6}$  &$-1$ &  $-\frac{10}{30} $  &$ \frac{3}{4}$&
$\frac{1}{4}     $
&44\\ \hline\hline
$D^+$ &$-1$ &$2$ & $\frac{13}{30}$  & $  0$&  $ -\frac{3}{4}$  & 63\\
\hline\hline
$D^0$ &$\frac{1}{2}$ &-1  &    $-\frac{10}{30}$ &   $  0$&
  $ -\frac{3}{4 } $ &39     \\
\hline\hline
 $D^{*+}$ &$\frac{1}{3}$ &$2$ & $\frac{13}{30}$  & $ 0$&  $ \frac{1}{4} $
  &
 24\\
\hline\hline
 $D^{*0}$ &$-\frac{1}{6}$ &-1  &    $-\frac{10}{30}$ &   $ 0$
&  $ \frac{1}{4}$ &9     \\
\hline\hline
 $\eta_c $& $\frac{1}{2}$  &$-1$ &  $-\frac{10}{30} $
&$ \frac{3}{4}$& $-\frac{3}{4}  $
&29 \\ \hline\hline
$\Psi $& $-\frac{1}{6}$  &$-1$ &  $-\frac{10}{30} $  &$ \frac{3}{4}$&
 $\frac{1}{4} $
&44\\ \hline\hline
 \end{tabular}

\vl

Table 4.21. \label{Mesoisospinsplit}  Various second order contributions to
meson
 masses. The notation $E_m=2C_m= \frac{D_m}{18}$ and $C_e= \frac{D_e}{9}$
(slightly
 different as compared with  the  baryonic case) have  been used.

\vm

It is convenient to write second order correction to meson mass as

\begin{eqnarray}
X&=& \Delta X (Coul)+\Delta X(magn)+ \Delta_0 +\Delta X^I +
\Delta^{spin}\nonumber\\
\
\Delta X^I &= &-\frac{1}{2}\Delta^I(I(I+1)-\frac{3}{2}) \nonumber\\
\Delta X^{spin}&=& \frac{1}{2}\Delta^{spin} (J(J+1)-\frac{3}{2})
\end{eqnarray}

\noindent $\Delta^0$ is constant shift coming from sea and color Coulombic
 interaction. $\Delta^I$ is a shift coming from the interaction between
same generation isospins and has
 same dependence on isospin as color hyper fine splitting on
$\Delta^{spin}$ on spin.

\vm

  The values of the parameters
$C_e,C_m,\Delta_0,\Delta^I,\Delta^{spin}$ for nonstrange quarks can be
 determined by comparing
predictions with the actual values of $X_{eff,n}$ and one obtains for
 nonstrange mesons the following
equations determining the parameter values.

\begin{eqnarray}
\Delta_0&=& \frac{1}{12}
( X_{eff}(K^+)+2X_{eff}(K^0)+3(X_{eff}(K^{*+})+2X_{eff}(K^{*0})  ) -
\frac{7}{90}\nonumber\\
&=&
\frac{1}{2}+\frac{1}{64\cdot 12}-\frac{7}{90}\nonumber\\
C_m(d,d)&=& \frac{3}{14}( X_{eff}(\pi^0)-X_{eff}(\rho^0)-
X_{eff}(\pi^+)+X_{eff}(\rho^+) )=  -\frac{3}{128} \nonumber\\
\Delta^{spin}(d,d)&=&  -X_{eff}(\pi^+)+X_{eff}(\rho^+)
-\frac{8C_m}{3}=\frac{1}{32}\nonumber\\
 C_e(d,d)&=& C_m(d,d)- \frac{2(X_{eff}(\pi^0)-X_{eff}(\pi^+)  )
}{9}=\frac{1}{9\cdot16}-\frac{3}{ 128}\nonumber\\
\Delta^I(d,d)&=&4(
-2C_m(d,d)+2C_e(d,d)-X_{eff}(\pi^+)+\frac{13}{30}-
\frac{3\Delta^{spin}(d,d)}{4}+\Delta_0 )\nonumber\\
&=&-\frac{7}{32}-\frac{1}{45}+\frac{1}{3\cdot 64}\nonumber\\
\
\end{eqnarray}

\noindent  The formulas are obtained from $X_{eff}(K^+) +2X_{eff}(K^0)+
3(X_{eff}(K^{+*}) + 2X_{eff}(K^{0*})$ , $\pi^0-\rho^0$ and $\pi^+-\rho^+$
mass differences and $\pi^+$ mass. The absolute value of  $C_e=
-19/(9\cdot 128)$ is considerably smaller  than its baryonic  counterpart
$C_e(d,d,B)\simeq 1/6$: a  possible interpretation is that pion radius is
about 10 times larger than proton radius.
 $C_m\simeq -3/128$ is much smaller   than  $C_m(d,d,B)\simeq 2/9$.  The
value of $C_e$ and $C_m$ are negative,
 which looks peculiar.   $C_m$ and $C_e$ are however  determined only
modulo integer (multiple of $6$
 for $C_m$)  and this multiple gives extremely small contribution to mass
squared. One cannot  exclude the possibility that all parameters are
nearly integers and that the the values deduced for them  correspond to the
deviations from integers.

\vm

One can predict the value of $X(\omega)$: \\
 $X(\omega)= X(\rho)-\Delta^I\simeq
\frac{29}{64}-\Delta^I=\frac{43}{64}+\frac{1}{45}+\frac{1}{3\cdot 64}$:\\
 The  real counterpart   $(pX(\omega))_R \simeq \frac{46}{64}$ is   not
far from
 the experimental value  $\frac{51}{64}$.    $X(\eta)$ can be predicted
reliably if the effects resulting from mixing with $\eta'$ and $\eta_c$
are negligible.  One gets  $X(\eta)=X(\pi^0)+\Delta^I\simeq
\frac{10}{64}-\frac{1}{45}+\frac{1}{3\cdot 64}$,
 whose real  is  certainly too too small as compared with the actual value
 $\frac{19}{64}$.   The real counter part of the second  order term is
sensitive to mixing effects so that the result is not a catastrophe.

\vm

For strange and charmed mesons the parameters $C_m(d,s)$,$C_e(d,s)$ and
$\Delta^{spin}(d,s)$ can be deduced from   $K^{^*}-K^+$, $K^{0*}-K^0$ and
$K^+$ mass assuming
 $\Delta^I$ to be negligibely small ($\Delta^I$ contributes to
$K^0(CP=1)-K^0(CP=-1)$ mass difference).
 Similar procedure can be used to deduce   $C_m(d,c)$,$C_e(d,c)$ and
$\Delta^{spin}(d,c)$.   In this
 manner one obtains

\begin{eqnarray}
C_m(d,s)&=& (X(K^{*+})-X(K^+))-(X(K^{*0})-X(K^0))=-\frac{5}{64}\nonumber\\
\Delta^{spin}(d,s)&=& \frac{1}{3}((X(K^{*+})-X(K^+))
+2(X(K^{*0})-X(K^0))=\frac{9}{16}+\frac{1}{192}\nonumber\\
C_e(d,s)&=&
C_m(d,s)+\frac{3}{8}\Delta^{spin}(d,s)+ \frac{X(K^+)}{2} -\frac{13}{60} +
\frac{\Delta_0}{2}\nonumber\\
&=&\frac{12}{64}+\frac{1}{30}- \frac{1}{192}\nonumber\\
\
\end{eqnarray}

\noindent   The value of $C_m(d,s)$ has same order of magnitude and sign
as  $C_m(d,d)$ but $C_e(d,s)$ and $\Delta^{spin}$ are much  larger than for
nonstrange mesons.

\vm

For charmed mesons $D^+,D^0,D^{*+},D^{*0}$  similar formulas apply and give
 the following values for the parameters

\begin{eqnarray}
C_m(d,c)&=& -\frac{9}{64}\nonumber\\
\Delta^{spin}(d,c)&=&\frac{7}{64}+\frac{1}{192}\nonumber\\
C_e(d,c)&=&\frac{25}{64}+\frac{1}{30}+ \frac{1}{6 \cdot 64}\nonumber\\
\
\end{eqnarray}

\noindent  (d,c)-elements are if same sign as (d,s)- elements but larger.

\vm

 The values of $X_{eff}$  for $\eta$, $\eta^,$, $\eta_c$, $\Phi$ and
$\Psi$ involve besides the known
 mixing angles the parameters $C_m(i,i),C_e(i,i)$,  $\Delta^{spin} (i,i)$,
$\Delta^I(i,i)$, $i=s,c$, 8
 parameters altogether,  so that masses can be reproduced with several
choices of parameter values  but
 predictions are not  possible.

\vm

In the previous calculations the assumption $\Delta
^I(d,s)=0,\Delta^I(d,c)=0$
 has been  made. The nonvanishing of this parameter induces splitting
between  $K^+,K^0, K^0(CP=-1)$  $I=1$  multiplet (same for $D$ mesons)
and $ K^0(CP=-1)$  $I=0$ multiplet and explains  the mass splitting of
neutral kaon system.
 The value of the splitting parameter  $\Delta^I(d,s)$   can be estimated:

\begin{eqnarray}
\Delta ^I(d,s)&=& \frac{2m_K\Delta m}{1024^2m_0^2}\simeq  \frac{2\delta m}
{m_K}s_{eff}(K)
\sim 10^{-13}\sim 2^{-44} \sim \frac{1}{\sqrt{M_{89}}}\nonumber\\
s_{eff}(K)&=&5
\end{eqnarray}

\noindent  Probably the appearence of $M_{89}$ is not an accident: the
decay
 of kaon to intermediate gauge boson pair explains mass splitting in
standard model and $M_{89}$ is
 the condensation level of intermediate gauge bosons.  It should be
noticed that the well known  $\Delta I=1/2$ rule becomes conservation law
for electroweak isospin in  TGD: for instance the decays of
 $K^0 (CP=-1)$ to two pions are strongly suppressed since initial state
corresponds to
 $I_{ew}=0$ state and final state has $I_{ew}=1$ by Bose statistics. The
observed  decays are made
 possible by  CP breaking implying a small  mixing of $I_{ew}=0$ and
$I_{ew}=1$ states.

 \section{ The observed top quark and $M_{89}$ physics?}

Top quark form the only exception in the nice general picture.
The TGD:eish predictions for the top mass  for $k=89$
 and $k=97$
levels and are
$m_t(89)\simeq 871 \ GeV $ and  $ m_t(97) \simeq 60.7 \ GeV $ to be compared
with the mass  $ m_t(obs)\simeq 174 \ GeV$ of the observed top candidate.
 The study of CKM matrix led to the cautious conclusion that only the
experimental  top candidate is consistent with CP breaking in
 $K-\bar{K}$ system so that observed top does not seem to correspond to
 neither
$k=97$ nor $k=89$ condensate level.   A possible
explanation  of the discrepancy is mixing of primary  condensate levels:
 observed
top  corresponds to the actual top, for which small mixing of condensate
 level
$k=97$ with condensate level $k=89$ takes place:

\begin{eqnarray}
t&=& cos(\Phi)t_{97}+sin(\Phi)t_{89}\nonumber\\
sin^2(\Phi) &=&\frac{ m_t^2-m_t^2(97)}{ m_t^2(89)-m_t^2(97)}\sim .036
\end{eqnarray}

\noindent The value of the mixing angle is rather small and means that
top spends
 less
than 4 per cent of its time on $k=89$ level.

\vl

\begin{center}
{\bf Acknowledgements\/}
\end{center}

\vm

It would not been possible to carry out this work without the  concrete
help of
 my friends  in concrete problems of the everyday life and I want to
express my gratitude to  them.  Also I want to thank J. Arponen,  R.
Kinnunen and J.  Maalampi
    for practical help and interesting discussions.

\newpage

\appendix{Mass fit for hadrons}

The tables below gives the  parameters $(s(H),n)$ for hadrons
in  the fit

\begin{eqnarray}
M^2 (H) &=&  (s(H)+\frac{n(H)}{64})M_0^2\frac{M_{127}}{M_{107}}\nonumber\\
M_0^2&=& \frac{m_e^2 }{5+\frac{2}{3}+\frac{1}{64}}\nonumber\\
M^2(H)&\simeq& (s+ \frac{n}{64}) M_0^2
\end{eqnarray}

\noindent  The approximate representation
 $M^2\simeq  (s+ \frac{n}{64} )M_0^2$ is suggested by the small quantum
number
 limit for Ramond representation and provides an   excellent fit of baryon
masses, the relative  errors being  below $2^{-11}$. For light mesons the
errors are few per cent.  For $\Delta^{-}$ one must have  $m\leq 1237.5 \
MeV$  instead of
 $m(\Delta^{++}+ (7.9 \pm 6.8)  \ MeV$ in order to obtain same  value of
$s(eff)$ for the entire multiplet. The
 precise measurement of $\Delta^-$ mass obviously gives a crucial test for
the scenario.

\vl

\begin{tabular}{||l|l|l|l|l||} \hline\hline
baryon &$ m_{exp}/MeV$ &s &$n$ &$10^3(\Delta M/M)$\\
\cline{1-5}\hline \hline
$p$ &938.2796&18 &25 &.13\\ \hline
$n$ &939.5731&18 & 28 &.82\\ \hline
$\Delta^{++}$&1231  &31 &41&.6\\ \hline
$\Delta^{+}$&1234.8 &31 &54&.9\\ \hline
$\Delta^{0}$&1233.6 &31 &50&.8\\ \hline
$\Delta^{-}$&$\leq 1237.5$ &31 &$\leq 63$&?\\ \hline
$\Lambda$ &1115.60   &25 &63&1.112\\ \hline
$\Sigma^+$ &1189.37  &29 &35&.69\\ \hline
$\Sigma^0$ &1192.37&29 &45&.88  \\ \hline\
$\Sigma^-$ &1197.35& 29&60&.99\\ \hline
$\Sigma^{*+}$ &1385  &40 &4&.076\\ \hline
$\Xi^{0}$ &1314.9& 36&7 &.12 \\ \hline
$\Xi^-$ &1321.29&36& 29&.32 \\ \hline
$\Xi^{*0}$ &1531.8 & 49&0 &-.029 \\ \hline
$\Xi^{*-}$ &1534.97&49&13&.11 \\ \hline
$\Omega^-$ &1672.2&58&25&.16\\ \hline
$\Lambda_c$ &2282.2&108&50&.27\\ \hline
$\Lambda_b$ &5425&614&41&.04\\ \hline\hline
\end{tabular}

\vl

Table 5.1. \label{Baryonmassfit} Mass fit for baryons.

\vl

\begin{tabular}{||l|l|l|l|l||}\hline\hline
meson &$ m_{exp}/MeV$   &s &X&$10^2(\Delta M/M)$\\\cline{1-5}\hline\hline
$\pi^0$ & 134.9645&0&24 &2.5\\ \hline
$\pi^+$ & 139.5688&0&26&3.1\\ \hline
$\rho^0$ & 772&12&29&.15\\ \hline
$\rho^+$ & 770&12&24&.07\\ \hline
$\omega$ & 783&12&51&.18\\ \hline
$\eta$ & 548.9&6&19&.2\\ \hline
$K^+$ & 493.707&5&6&.094\\ \hline
$K^0$ & 497.7&5&11&.099\\ \hline
$K^{*+}$ & 891.77&16&39&.13\\ \hline
$K^{*0}$ & 896.05&16&49&.14\\ \hline
$\eta'$ & 957.6&19&10&.04\\ \hline
$\Phi$ & 1019&21&44&.1\\ \hline
$D^+$ &1869.4 &  72&63&.05\\ \hline
$D^0$ &1864.7 &  72&39&.02\\ \hline
$D^{*+}$& 2010.1& 84&24&.01 \\ \hline
$D^{*0}$& 2007.2& 84&9 &.006\\ \hline
$F$& 2021 &85&19&.0097\\ \hline
$\eta_c$&2980&185&29&.006\\ \hline
$\Psi$&3100&200&45&.01\\ \hline
$B$&5270&580&1&.0003\\\hline
$Y$&9460& 1868&61&.002\\ \hline\hline
\end{tabular}

\vl

Table 5.2. \label{Mesonmassfit} Mass fit for mesons.
The relative errors for pions are order 3  per  cent.

\end{document}